\documentclass[10pt,journal,compsoc]{IEEEtran}
\usepackage{amsfonts}
\usepackage{algorithm}
\usepackage{algorithmic}
\usepackage{amsmath}
\usepackage{amssymb}
\usepackage{amsfonts}
\usepackage{bm}
\usepackage{color}
\usepackage{comment}
\usepackage{graphicx}
\usepackage{multirow}
\usepackage{subfigure}
\usepackage{url}

\newtheorem{definition}{Definition}

\ifCLASSOPTIONcompsoc

\else

\fi

\ifCLASSINFOpdf

\else

\fi

\begin{document}

\title{ DiffNet++: A Neural Influence and Interest Diffusion Network for Social Recommendation}

\author{Le Wu,~\IEEEmembership{Member,~IEEE}, Junwei Li, Peijie Sun,  Richang Hong,~\IEEEmembership{Member,~IEEE},
 \\
Yong Ge,~\IEEEmembership{Member,~IEEE}, Meng Wang,~\IEEEmembership{Fellow,~IEEE}

\IEEEcompsocitemizethanks{

\IEEEcompsocthanksitem  L.~Wu, M.~Wang are with the School of Computer Science and Information Engineering, Hefei University of Technology,
Hefei, Anhui 230009, China, and Institute of Artificial Intelligence, Hefei Comprehensive National Science Center, Hefei, Anhui 230088. \protect \\Emails: \{lewu.ustc, eric.mengwang\}@gmail.com.

\IEEEcompsocthanksitem  J. Li, P.~Sun, R.~Hong~(corresponding author) are with the School of Computer Science and Information Engineering, Hefei University of Technology,
Hefei, Anhui 230009, China.
\protect \\Emails: \{lijunwei.edu, sun.hfut\}@gmail.com.

\IEEEcompsocthanksitem Y.~Ge is with Management Information Systems Department, The University of
Arizona, Tucson, Arizona, USA. \protect Email:yongge@email.arizona.edu.

}
}

\markboth{IEEE TRANSACTIONS ON KNOWLEDGE AND DATA ENGINEERING}
{Shell \MakeLowercase{\textit{et al.}}: Bare Advanced Demo of IEEEtran.cls for Journals}

\IEEEtitleabstractindextext{
\begin{abstract}

Social recommendation has emerged to leverage social connections among users for predicting users' unknown preferences, which could alleviate the data sparsity issue in collaborative filtering based recommendation.
Early approaches relied on utilizing each user's first-order social neighbors' interests for better user modeling, and failed to model the social influence diffusion process from the global social network structure.  Recently, we propose a preliminary work of a neural influence \emph{Diff}usion \emph{Net}work~(i.e., DiffNet) for social recommendation~\cite{leDiffnet}. DiffNet models the recursive social diffusion process for each user, such that the influence diffusion hidden in the higher-order social network is captured in the user embedding process. Despite the superior performance of DiffNet, we argue that, as users play a central role in both user-user social network and user-item interest network, only modeling the influence diffusion process in the social network would neglect the latent collaborative interests of users hidden in the user-item interest network. To this end, in this paper, we propose DiffNet++, an improved algorithm of DiffNet that models the neural influence diffusion and interest diffusion in a unified framework. By reformulating the social recommendation as a heterogeneous graph with social network and interest network as input, DiffNet++ advances DiffNet by injecting both the higher-order user latent interest reflected in the user-item graph and higher-order user influence reflected in the user-user graph for user embedding learning. This is achieved by iteratively aggregating each user's embedding from three aspects: the user's previous embedding, the influence aggregation of social neighbors from the social network, and the interest aggregation of item neighbors from the user-item interest network. Furthermore, we design a multi-level attention network that learns how to attentively aggregate user embeddings from these three aspects. Finally, extensive experimental results on four real-world datasets clearly show the effectiveness of our proposed model. We release the source code at  
\url{https://github.com/PeiJieSun/diffnet}.

\end{abstract}


}

\maketitle


\IEEEpeerreviewmaketitle

\ifCLASSOPTIONcompsoc
\else

\fi

\section{Introduction}

Collaborative Filtering~(CF) based recommender systems learn user and item embeddings by utilizing user-item interest behavior data, and have attracted attention from both the academia and industry~\cite{bpr,pmf}. However, as most users have limited  behavior data, CF suffers from the data sparsity issue~\cite{adomavicius2005toward}.  With the development of social networks, users build social relationships and share their item preferences on these platforms. As well supported by the social influence theory, users in a social network would influence each other, leading to similar preferences~\cite{friedkin2006structural,KDD2008influence}. Therefore, social recommendation has emerged, which focuses on exploiting social relations among users to alleviate data sparsity and enhancing recommendation performance~\cite{socialmf,jiang2012social,trustsvd,leDiffnet}.

In fact, as users play a central role in social platforms with user-user social behavior and user-item interest behavior, the key to social recommendation relies on learning user embeddings with these two kinds of behaviors. For a long time, by treating the user-item interest network as a user-item matrix, CF based models resort to matrix factorization to project both users and items into a low latent space~\cite{bpr,pmf,rendle2010factorization}.
Most social based recommender systems advance these CF models by leveraging the user-user matrix to enhance each user's embedding learning with social neighbors' records, or regularizing the user embedding learning process with social neighbors~\cite{socialmf,WSDM2011recommender, tkde2014scalable,TKDE2016novel}. For example, SocialMF~\cite{socialmf} and SR~\cite{WSDM2011recommender} added social regularization terms based on social neighbors in the optimization function, and TrustSVD incorporated influences of social neighbors' decisions as additional terms for modeling a user's embedding~\cite{TKDE2016novel}. In summary, these models leveraged the first-order social neighbors for recommendation, and partially alleviated the data sparsity issue in CF.

\begin{small}
\begin{figure*} [htb]
 \begin{center}
 \vspace{-0.3cm}
\includegraphics[width=160mm]{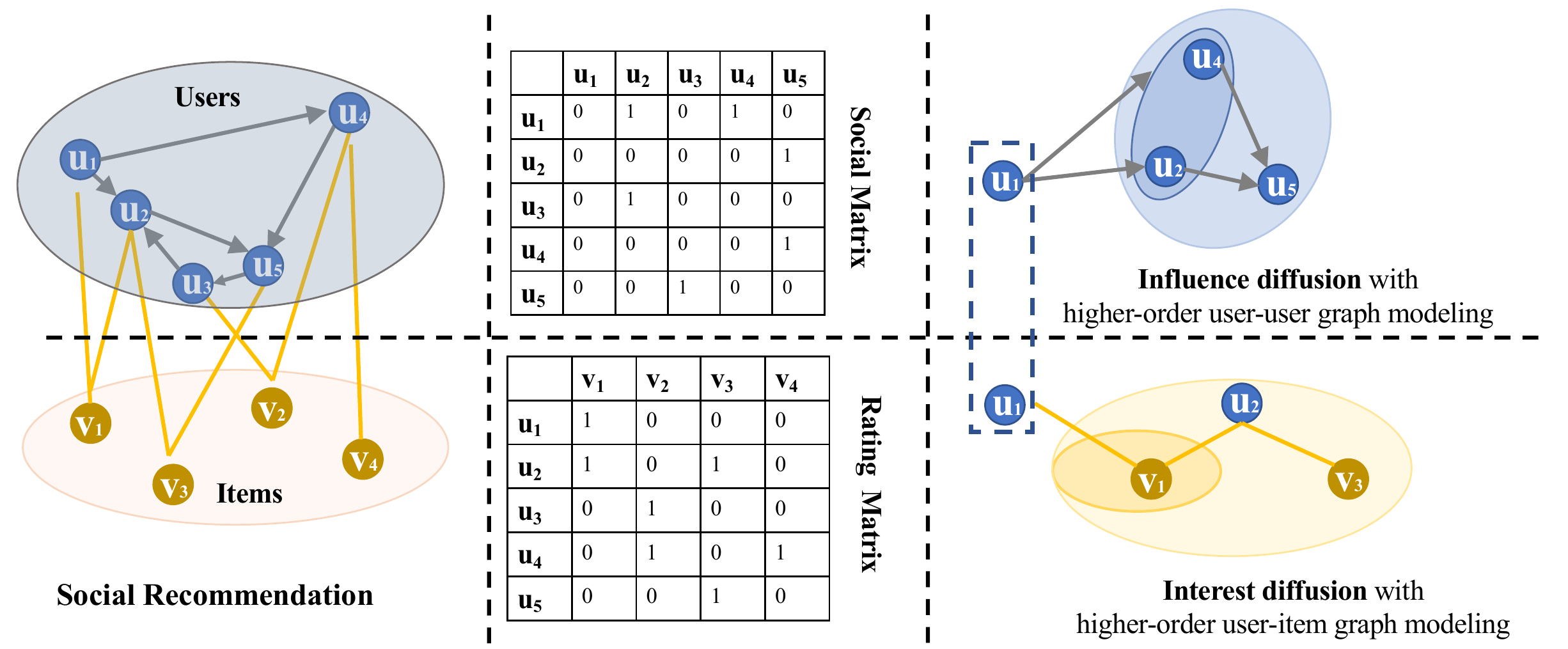}
  \end{center}
  \vspace{-0.2cm}
  \caption{\small{An overall illustration of social recommendation. The second column shows how traditional models
  treat this problem with matrix representations of users' two kinds of behaviors. In this paper, we try to model both the influence diffusion and interest diffusion with graph representation of users' two kinds of behaviors. } }
  \vspace{-0.4cm}
\label{fig:example_intro}
\end{figure*}
\end{small}

Despite the performance improvement of these social recommendation models, we argue that the current social recommendation models are still far from satisfactory. In fact, as shown in Fig.~\ref{fig:example_intro}, users play a central role in two kinds of behavior networks: the user-user social network and the user-item interest network. On one hand, users naturally form a social graph with a global recursive social diffusion process. Each user is not only influenced by the direct first-order social neighbors, but also the higher-order ego-centric social network structure. E.g., though user $u1$ does not follow $u5$,  $u1$ may be largely influenced by $u5$ in the social recommendation process as there are two second-order paths: $u1\rightarrow u2\rightarrow u5$ and $u1\rightarrow u4\rightarrow u5$. Simply reducing the social network structure to the first-order social neighbors would not well capture these higher-order social influence effect in the recommendation process. On the other hand, given the user-item bipartite interest graph, CF relies on the assumption that  ``similar users show similar item interests''.  Therefore, each user's latent collaborative interests are not only reflected by her rated items but also influenced by similar users' interests from items. E.g, though $u1$ does not show interests for $v3$ with a direct edge connection, the similar user $u2$ (as they have common item interests of $v1$) shows item interest for $v3$ as: $u1\leftrightarrow v1\leftrightarrow u2\leftrightarrow v3$. Therefore, $v3$ is also useful for learning $u1$'s embedding to ensure the collaborative signals hidden in the user-item graph are injected for user embedding learning. To summarize, previous CF and social recommendation models only considered the observed first-order structure of the two graphs, leaving the higher-order structures of users under explored.

To this end, we reformulate users' two kinds of behaviors as a heterogeneous network with two graphs, i.e, a user-user social graph and  a user-item interest graph, and propose how to explore the heterogeneous graph structure for social recommendation. In fact, Graph Convolutional Networks~(GCNs) have shown huge success for learning graph structures with theoretical elegance, practical flexibility and high performance~\cite{bruna2013spectral,defferrard2016convolutional,kipf2016semi}. GCNs perform node feature propagation in the graph, which recursively propagate node features by iteratively convolutional aggregations from neighborhood nodes, such that the up to $K$-th order graph structure is captured with $K$ iterations~\cite{ICLR2019powerful}.  By treating user-item interactions as a bipartite interest graph and user-user social network as a social graph, some works have applied GCNs separately on these two kinds of graphs~\cite{zheng2018spectral,wang2019neural,ying2018graph,leDiffnet}. On one hand, given the user-item interest graph, NGCF is proposed to directly encode the collaborative information of users by exploring the higher-order connectivity patterns with embedding propagation~\cite{wang2019neural}. On the other hand, in our previous work, we propose a \emph{Diff}usion neural \emph{Net}work~(DiffNet) to model the recursive social diffusion process in the social network, such that the higher-order social structure is directly modeled in the recursive user embedding process~\cite{leDiffnet}. These graph based models showed superior performance compared to the previous non-graph based recommendation models by modeling either graph structure. Nevertheless,  how to design a unified model for better user modeling of these two graphs remains under explored.

In this paper, we propose to advance our preliminary DiffNet structure, and jointly model the two graph structure~(user-item graph and user-user graph) for social recommendation. While it seems intuitive to perform message passing on both each user's social network and interest network, it is not well designed in practice as these two kinds of graphs serve as different sources to reflect each user's latent preferences. Besides, different users may have different preferences in balancing these two graphs, with some users are likely to be swayed by social neighbors, while others prefer to remain their own tastes. To this end,  we propose DiffNet++, an improved algorithm of DiffNet that models the neural influence diffusion and interest diffusion in a unified framework. Furthermore, we design a multi-level attention network structure that learns how to attentively aggregate user embeddings from different nodes in a graph, and then from different graphs. In summary, our main contributions are listed as follows:

\begin{itemize}
\item Compared to our previous work of DiffNet~\cite{leDiffnet}, we revisit the social recommendation problem as predicting the missing edges in the user-item interest graph by taking both user-item interest graph and user-user social graph as input.

\item We propose DiffNet++ that models both the higher-order social influence diffusion in the social network and interest diffusion in the interest network in a unified model. Besides,  we carefully design a multi-level attention network to attentively learn how the users prefer different graph sources.

\item Extensive experimental results on two real-world datasets clearly show the effectiveness of our proposed DiffNet++ model. Compared to the baseline with the best performance, DiffNet++ outperforms it about 14\% on Yelp , 21\% on Flickr, 12\% on Epinions, and 4\% on Dianping for top-10 recommendation.
\end{itemize}

\section{Problem Definition and Related Work}

\subsection{Problem Definition}
In a social recommender system, there are two sets of entities: a user set {\small$U$~($|U|\!=\!M$)}, and an item set {\small$V$~($|V|\!=\!N$)}. Users form two kinds of behaviors in the social platforms: making social connections with other users and showing item interests. These two kinds of behaviors could be defined as two matrices: a user-user social connection matrix  {\small $\mathbf{S}\in\mathbb{R}^{M\times M}$}, and a user-item interaction matrix  {\small$\mathbf{R}\in\mathbb{R}^{M\times N}$}. In the social matrix {\small$\mathbf{S}$}, if user $a$ trusts or follows user $b$, $s_{ba}\!=\!1$, otherwise it equals 0. We use $S_a$ to represent the user set that user $a$ follows, i.e., $S_a\!=\![b|s_{ba}\!=\!1]$. The user-item matrix {\small$\mathbf{R}$} shows users' rating preferences and interests to items. As some implicit feedbacks~(e.g., watching movies, purchasing items, listening to songs ) are more common in real-world applications, we also consider the recommendation scenario with implicit feedback~\cite{bpr}. Let {\small $\mathbf{R}$} denote users' implicit feedback based rating matrix, with $r_{ai}\!=\!1$ if user $a$ is interested in item $i$, otherwise it equals 0. We use $R_a$ represents the item set that user $a$ has consumed, i.e., $R_a\!=\![i|r_{ai}\!=\!1]$, and $R_i$ denotes the user set which consumed the item $i$, i.e., $R_i\!=\![a|r_{ia}\!=\!1]$.

Given the two kinds of users' behaviors, the user-user social network is denoted as a user-user directed graph: {\small$G_S = <U, \mathbf{S}>$}, where $U$ is the nodes of all users in the social network. If the social network is undirected, then user $a$ connects to user $b$ denotes $a$ follows $b$, and $b$ also follows $a$, i.e., $s_{ab}\!=\!1\!\wedge s_{ba}\!=\!1$. The user interest network denotes users' interests for items,  which could be constructed from the user-item rating matrix {\small$\mathbf{R}$} as an undirected bipartite network: {\small$G_I = <U\cup V, \mathbf{R}>$}.

Besides, each user $a$ is associated with real-valued attributes~(e.g, user profile), denoted as $\mathbf{x}_a$ in the user attribute matrix {\small $\mathbf{X}\in\mathbb{R}^{d1\times M}$}. Also, each item $i$ has an attribute vector $\mathbf{y}_i$~(e.g., item text representation, item visual representation) in item attribute matrix {\small $\mathbf{Y}\in\mathbb{R}^{d2\times N}$}.  We formulate the graph based social recommendation problem as:

\begin{definition}  [{\small \textbf{Graph Based Social Recommendation}}]Given the user social network $G_S$ and user interest network {\small$G_I$}, these two networks could be formulated as a heterogeneous graph that combines {\small$G_S$} and {\small$G_I$} as: {\small$G=G_S\cup G_I=<U\cup V, \mathbf{X}, \mathbf{Y}, \mathbf{R}, \mathbf{S}>$}. Then, the graph based social recommendation asks that: given graph $G$ in the social network, our goal is to predict users' unknown preferences to items , i.e, the missing links in the graph based social recommendation as : $\hat{R}=f(G)=f(U\cup V, \mathbf{X}, \mathbf{Y}, \mathbf{R}, \mathbf{S})$, where {\small$\hat{R}\in\mathbb{R}^{M\times N}$} denotes the predicted preferences of users to items.
\end{definition}

\subsection{Preliminaries and Related Work}
In this subsection, we summarize the related works for social recommendation into three categories: classical social recommendation models, the recent graph based recommendation models, and attention modeling in the recommendation domain.

\textbf{Classical Social Recommendation Models.} By formulating users' historical behavior as a user-item interaction matrix {\small$\mathbf{R}$,  most classical CF models embed both users and items in a low dimension latent space, such that each user's predicted preference to an unknown item turns to the inner product between the corresponding user and item embeddings as~\cite{bpr,pmf,rendle2010factorization}:

\vspace{-0.2cm}
 \begin{equation}\label{eq:pred_r}
\hat{r}_{ai}=\mathbf{v}^T_i\mathbf{u}_a,
\end{equation}

\noindent where $\mathbf{u}_a$ is the embedding of user $a$, which is the a-th column of  the user embedding matrix {\small$\mathbf{U}$}. Similarly, $\mathbf{v}_i$} represents item $i$'s embedding in the $i$-th column of item embedding matrix {\small$\mathbf{V}$}.

In fact, as various specialized matrix factorization models have been proposed for specific tasks, factorization machines is proposed as a general approach to mimic most factorization models with simple feature engineering~\cite{rendle2010factorization}.  Recently, some deep learning based models have been proposed to tackle the CF problem~\cite{nmf,kdd2018xdeepfm}.  These approaches advanced previous works by modeling the non-linear complex interactions between users, or the complex interactions between sparse feature input.

The social influence and social correlation among users' interests are the foundation for building social recommender systems\cite{liu2014influence,sun2018attentive,PNAS2012social,PNAS2014experimental}. Therefore, the social network among users could be leveraged to alleviate the sparsity in CF and enhance recommendation performance~\cite{WSDM2011recommender,sun2018attentive,TKDE2016novel}. Due to the superiority of embedding based models for recommendation, most social recommendation models are also built on these embedding models. These social embedding models could be summarized into the following two categories: the social regularization based approaches~\cite{socialmf,WSDM2011recommender,tkde2014scalable,li2015overlapping,wu2018collaborative} and the user behavior enhancement based approaches~\cite{trustsvd,TKDE2016novel}. Specifically, the social regularization based approaches assumed that connected users would  show similar embeddings under the social influence diffusion. As such, besides the classical CF pair-wised loss function in BPR~\cite{bpr}, an additional social regularization term is incorporated in the overall optimization function as:

\begin{small}
\vspace{-0.2cm}
\begin{equation}
\sum_{i=1}^M\sum_{j=1}^M s_{ij}||\mathbf{u}_i-\mathbf{u}_j||_F^2=\mathbf{U}(\mathbf{D}-\mathbf{S})\mathbf{U}^T,
\end{equation}
\vspace{-0.2cm}
\end{small}

\noindent where {\small $\mathbf{D}$} is a diagonal matrix with  $d_{aa}=\sum_{b=1}^M s_{ab}$.

Instead of the social regularization term, some researchers argued that the social network provides valuable information to enhance each user's behavior~\cite{zhao2014leveraging,TKDE2016novel}. TrustSVD is such a representative model that shows state-of-the-art performance~\cite{trustsvd,TKDE2016novel}. By assuming the implicit feedbacks of a user's social neighbors' on items could be regarded as the auxiliary feedback of this user, TrustSVD modeled the predicted preference as:

\vspace{-0.2cm}
\begin{equation}\label{eq:trustsvd}
\hat{r}_{ai}=\mathbf{v}^T_i (\mathbf{u}_a+|R_a|^{\frac{-1}{2}}\sum_{i\in R_a}\mathbf{y}_i +|S_a|^{-\frac{1}{2}}\sum_{b\in S_a}\mathbf{u}_b)
\end{equation}
\vspace{-0.2cm}

\noindent where {\small$R_a\!=\![i|r_{ai}\!=\!1]$} is the itemset that $a$ shows implicit feedback, and $\mathbf{y}_i$ is an implicit factor vector. Therefore, these first two terms compose SVD++ model that explicitly builds each user's liked items in the user embedding learning process~\cite{koren2008factorization}. In the third term, $\mathbf{u}_b$ denotes the latent embedding of user $b$, who is trusted by $a$. As such, $a$'s latent embedding is enhanced by considering the influence of her trusted users' latent embeddings in the social network.

As items are associated with attribute information~(e.g., item description, item visual information), ContextMF is proposed to combine social context and social network under a collective matrix factorization framework with carefully designed regularization terms~\cite{tkde2014scalable}. Social recommendation has also been extended with social circles~\cite{tkde2014personalized}, temporal context~\cite{sun2018attentive},  rich contextual information~\cite{wu2019hierarchical}, user role in the social network~\cite{tang2013exploiting}, and efficient training models without negative sampling~\cite{SIGIR2019efficient}. All these previous works focused on how to explore the social neighbors, i.e., the observed links in the social network. Recently, CNSR is proposed to leverage the global social network in the recommendation process~\cite{wu2018collaborative}. In CNSR, each user's latent embedding is composed of two parts: a free latent embedding~(classical CF models), and a social network embedding that captures the global social network structure. Despite the relative improvement of CNSR, we argue that CNSR is still suboptimal as the global social network embedding process is modeled for the network based optimization tasks instead of user preference learning. In contrast to CNSR, our work explicitly models the recursive social diffusion process in the global social network for optimizing the recommendation task. Researchers proposed to generate social sequences based on random walks on user-user and user-item graph, and further leveraged the sequence embedding techniques for social recommendation~\cite{fan2019deep}. This model could better capture the higher-order social network structure. However, the performance heavily relies on the choice of random walk strategy, including switching between user-item graph and user-user graph, and the length of random walk, which is both time-consuming and labor-consuming.

\textbf{Graph Convolutional Networks and Applications in Recommendation.}
GCNs generalize the convolutional operations from the regular Euclidean domains to non-Euclidean graph and have empirically shown great success in graph representation learning~\cite{bruna2013spectral,defferrard2016convolutional,kipf2016semi}. Specifically, GCNs recursively perform message passing by applying convolutional operations to aggregate the neighborhood information, such that the $K$-th order graph structure is captured with $K$ iterations~\cite{kipf2016semi}. By treating the user-item interaction as a graph structure, GCNs have been applied for recommendation~\cite{ying2018graph,zheng2018spectral}.
Earlier works relied on spectral GCNs, and suffered from huge time complexity~\cite{monti2017geometric,zheng2018spectral}. Therefore, many recent works focus on the spatial based GCNs for recommendation~\cite{ying2018graph,arXiv2017gcmc,ijcai2019star-gcn,wang2019neural}. PinSage is a GCN based content recommendation model by propagating item features in the item-item correlation graph~\cite{ying2018graph}.  GC-MC applied  graph neural network for CF, with the first order neighborhood is directly modeled in the process~\cite{arXiv2017gcmc}. NGCF extended GC-MC with multiple layers, such that the higher-order collaborative signals between users and items can be modeled in the user and item embedding learning process~\cite{wang2019neural}.

As the social structure among users could be naturally formulated as a user-user graph, recently we propose a preliminary graph based social recommendation model, DiffNet, for modeling the social diffusion process in recommendation~\cite{leDiffnet}. DiffNet advances classical embedding based models with carefully designed influence diffusion layers, such that how users are influenced by the recursive influence diffusion process in the social network could be well modeled. Given each user $a$, the user embedding  $\mathbf{u}_{a}$ is sent to the influence diffusion layers.  Specifically, let $K$ denote the depth of the influence diffusion layers and  $\mathbf{h}^k_a$
is the user representation at the $k$-th layer of the influence diffusion part. For each user $a$,  her updated embedding $\mathbf{h}^{k+1}_a$ is performed by social diffusion of the embeddings at the $k$-th layer with two steps: aggregation from her social neighbors at the $k$-th layer~(Eq.\eqref{eq:diffnet_agg}), and combination of her own latent embedding $\mathbf{h}^k_a$ at $k$-th layer and neighbors:

\begin{eqnarray}\label{eq:diffnet}
\mathbf{h}^{k+1}_{Sa}=Pool(\mathbf{h}^k_b| b\in S_a), \label{eq:diffnet_agg} \\
\mathbf{h}^{k+1}_a=s^{(k+1)}(\mathbf{W}^k\times[\mathbf{h}^{k+1}_{S_a}, \mathbf{h}^k_a]),
\end{eqnarray}

\noindent where the first equation is a pooling operation that transforms all the social trusted users’ influences into a fixed length vector $\mathbf{h}^{k+1}_{Sa}$, $s(x)$ is a transformation function and we use $s^{(k+1)}$ to denote the transformation function for $(k\!+\!1)$-th layer. As such, with a diffusion depth $K$, DiffNet could automatically models how users are influenced by the K-th order social neighbors in a social network for social recommendation. When {\small$K\!=\!0$}, the social diffusion layers disappear and DiffNet degenerates to classical CF models.

In summary, all these previous GCN based models either considered the higher-order social network or the higher-order user interest network for recommendation.  There are some recently works that also leverage the graph neural networks for social recommendation~\cite{fan2019graph,wu2019dual}. Specifically, GraphRec is designed to learn user representations by fusing first order social and first-order item neighbors with non-linear neural networks~\cite{fan2019graph}. Researchers also proposed deep learning techniques to model the complex interaction of dynamic and static patterns reflected from users' social behavior and item preferences~\cite{wu2019dual}. Although these works relied on deep learning based models with users' two kinds of behaviors, they only modeled the first order structure of the social graph and interest graph. We differ from these works as we simultaneously fuse the higher-order social and interest network structure for better social recommendation.

\textbf{Attention Models and Applications.} As a powerful and common technique, attention mechanism is often adopted when multiple elements in a sequence or set would have an impact of the following output, such that attentive weights are learned with deep neural networks to distinguish important elements~\cite{itti1998model,bahdanau2014neural,xu2015show}. Given a user's rated item history, NAIS  is proposed to learn the neural attentive weights for item similarity in item based collaborative filtering~\cite{he2018nais}.
For graph structure data, researchers proposed graph attention networks to attentively learn weights of each neighbor node in the graph convolutional process~\cite{velivckovic2017graph}.
In social recommendation, many attention models have been proposed to learn the social influence strength~\cite{qiu2018deepinf,sun2018attentive,gong2018adaptive,fan2019graph}. E.g., with each user's direct item neighbors and social neighbors, GraphRec leverages attention modeling to learn the attentive weights for each social neighbor and each rated item for user modeling~\cite{fan2019graph}. In  social contextual recommender systems, users' preferences are influenced by various social contextual aspect, and an attention network was proposed to learn the attention weight of each social contextual aspect in the user decision process. Our work is also inspired by the applications of attention modeling, and apply it to fuse the social network and interest network for social recommendation.

\section{The Proposed Model}
\begin{figure*}[!t]\label{model_DiffNet_plus}
\centering
\includegraphics[scale=0.6]{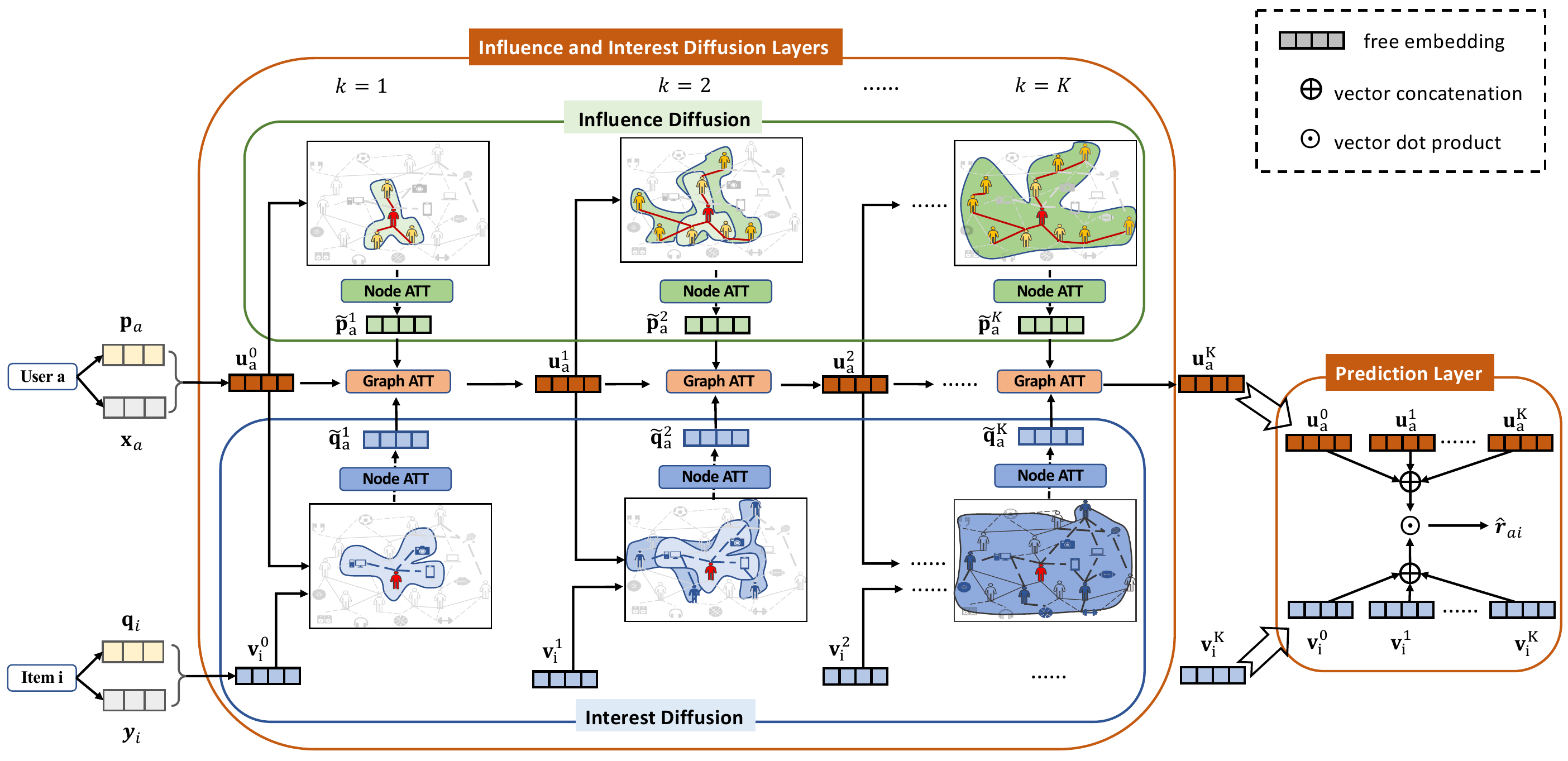}
\caption{The overall structure of the DiffNet++ model. As shown in the graph, we use \emph{Node ATT} to denote the node level attention layer in each graph, and \emph{Graph ATT} to denote the graph attention layer when fusing the interest graph representation and social graph representation.}
\label{model_DiffNet_plus}
\end{figure*}

In this section, we first show the overall architecture of our proposed model DiffNet++, followed by each component. After that, we will introduce the learning process of DiffNet++. Finally, we give a detailed discussion of the proposed model.

\subsection{Model Architecture}
As shown in the related work part, our preliminary work of DiffNet adopts the recursive influence diffusion process for iterative user embedding learning, such that the up to $K$-th order social network structure is injected into the social recommendation process~\cite{leDiffnet}. In this part,  we propose DiffNet++, an enhanced model of DiffNet that fuses both influence diffusion in the social network $G_S$ and interest diffusion in the interest network  $G_I$ for social recommendation. We show the overall neural architecture of DiffNet++ in Fig.~\ref{model_DiffNet_plus}. The architecture of DiffNet++ contains four  main parts: an embedding layer, a fusion layer, the influence and interest diffusion layers, and a rating prediction layer. Specifically, by taking related inputs, the embedding layer outputs free embeddings of users and items, and the fusion layer fuses both the content features and free embeddings. In the influence and interest diffusion layers, we carefully design a multi-level attention structure that could effectively
diffuse higher-order social and interest networks. After the diffusion process reaches stable, the output layer predicts the preference score of each unobserved user-item pair.

\textbf{Embedding Layer.} It encodes users and items with corresponding free vector representations. Let $\mathbf{P}\in \mathbb{R}^{ M\times D}$ and $\mathbf{Q}\in \mathbb{R}^{ N\times D}$ represent the free latent embedding matrices of users and items with $D$ dimensions. Given the one hot representations of user $a$, the embedding layer performs an index selection and outputs the free user latent embedding $\mathbf{p}_a$, i.e., the transpose of $a$-th row
from  user free embedding matrix $\mathbf{P}$. Similarly, item $i$'s embedding $\mathbf{q}_i$ is the transpose of $i$-th row of item free embedding matrix $\mathbf{Q}$.

\textbf{Fusion Layer.} For each user $a$, the fusion layer takes $\mathbf{p}_a$ and her associated feature vector $\mathbf{x}_a$ as input, and outputs a user fusion embedding $\mathbf{u}^0_a$ that captures the user's initial interests from different kinds of input data. We model the fusion layer as:

\begin{small}
\begin{equation}\label{eq:fusion_user}
\mathbf{u}^0_a = g(\mathbf{W}_1\times[\mathbf{p}_a,\mathbf{x}_a,]),
\end{equation}
\end{small}

\noindent where $\mathbf{W}_1$  is a transformation matrix, and $g(x)$ is a transformation function.  Without confusion, we omit the bias term. This fusion layer could generalize many typical fusion operations, such as the concatenation operation $\mathbf{u}^0_a=[\mathbf{p}_a, \mathbf{x}_a]$  by setting {$\mathbf{W}_1$} as an identity matrix and $g(x)$ an identity function.

Similarly, for each item $i$, the fusion layer models the item embedding  $\mathbf{v}_{i}^{0}$ as a function between its free latent vector $\mathbf{q}_i$ and its feature vector $\mathbf{y}_i$ as:

\vspace{-0.2cm}
\begin{equation} \label{eq:item_embed}
\mathbf{v}^0_i = g(\mathbf{W}_2\times[\mathbf{q}_i, \mathbf{y}_i]).
\end{equation}
\vspace{-0.2cm}

\textbf{Influence and Interest Diffusion Layers.} By feeding the output of each user $a$'s fused embedding $\mathbf{u}^0_a$ and each item $i$'s fused embedding $\mathbf{v}^0_i$ into the influence and interest diffusion layers, these layers recursively model the dynamics of this user's latent preference and  the item's latent preference propagation in the graph $\mathcal{G}$ with layer-wise convolutions. In detail, at each layer $k+1$, by taking user $a$'s embedding $\mathbf{u}^k_a$ and item $i$'s embedding $\mathbf{v}^k_i$ from previous layer $k$ as input, these layers recursively output the updated embeddings of $\mathbf{v}^{k+1}_i$ and $\mathbf{u}^{k+1}_a$ with diffusion operations. This iteration step starts at $k=0$ and stops when the recursive process reaches a pre-defined depth {\small$K$}. As each item only appears in the user-item interest graph $G_I$, in the following, we would first introduce how to update item embeddings, followed by the user embedding with influence and interest diffusions.

For each item $i$, given its $k$-th layer embedding $\mathbf{v}^k_i$,  we model the updated item embedding $\mathbf{v}^{k+1}_i$ at the $(k\!+\!1)$-th layer from $G_I$ as:

\begin{small}
\vspace{-0.2cm}
\begin{flalign}
\label{eq:item_agg}&  \tilde{\mathbf{v}}^{k+1}_i =AGG_u(\mathbf{u}^k_{a}, \forall a\in R_i)=\sum_{a\in R_i}\eta^{k+1}_{ia}\mathbf{u}^k_{a},\\
\label{eq:item_update} &\mathbf{v}^{k+1}_i=\tilde{\mathbf{v}}^{k+1}_i + \mathbf{v}^k_i,
\end{flalign}
\vspace{-0.4cm}
\end{small}

\noindent where $R_i=[a|r_{ia}=1]$ is the userset that rates item $i$. $\mathbf{u}^k_a$ is the $k$-th layer embedding of user $a$. $\tilde{\mathbf{v}}^{k+1}_{i}$ is the item $i$'s aggregated embedding from its neighbor users in the user-item interest graph $G_I$, with $\eta^{k+1}_{ia}$ denotes the aggregation weight.  After obtaining the aggregated embedding $\tilde{\mathbf{v}}^{k+1}_{i}$ from the $k$-th layer, each item's updated embedding $\mathbf{v}^{k+1}_i$ is a fusion of the aggregated neighbors' embeddings and the item's emebedding at previous layer $k$. In fact, we try different kinds of fusion functions, including the concatenation and the addition, and find the addition always shows the best performance. Therefore, we use the addition as the fusion function in Eq.\eqref{eq:item_update}.

In the item neighbor aggregation function, Eq.\eqref{eq:item_agg} shows the weight of user $a$ to item $i$. A naive idea is to aggregate the embeddings from $i$'s neighbor users with mean pooling
operation, i.e., $\tilde{\mathbf{v}}^{k+1}_i$=$\sum_{a\in R_i}\frac{1}{|R_i|}\mathbf{u}^k_{a}$. However, it neglects the different interest weights from users, as the importance values of different users vary in item representation. Therefore, we use an attention network to learn the attentive weight  $\eta^{k+1}_{ia}$ in Eq.\eqref{eq:item_agg} as:

\vspace{5pt}

\begin{small}
\vspace{-0.2cm}
\begin{equation} \label{eq:item_agg_mlp}
\eta^{k+1}_{ia} = MLP_1([\mathbf{v}^k_i, \mathbf{u}^k_a]),
\end{equation}
\vspace{-0.4cm}
\end{small}

\vspace{5pt}

\noindent where a MultiLayer Perceptrion~(MLP) is used to learn the node attention weights with the related user and item embeddings at the $k$-th layer. After that, we normalize the attention weights with:

\vspace{5pt}
\begin{small}
\vspace{-0.2cm}
\begin{equation}\label{eq:item_agg_normalize}
\eta^{k+1}_{ia}=\frac{exp(\eta^{k+1}_{ia})}{\sum_{b\in R_i}exp(\eta^{k+1}_{ib})}.
\end{equation}
\vspace{-0.4cm}
\end{small}
\vspace{10pt}

\noindent Specifically, the exponential function is used to ensure each attention weight is larger than 0.

For each user $a$, let $\mathbf{u}^k_a$ denote her latent embedding at the $k$-th layer. As users play a central role in both the social network $G_S$ and the interest network $G_I$, besides her own latent embedding $\mathbf{u}^k_a$, her updated embedding $\mathbf{u}^{(k+1)}_a$ at $(k+1)$-th layer is influenced by two graphs:  the influence diffusion in $G_S$ and the interest diffusion in $G_I$.  Let
$\tilde{\mathbf{p}}^{k+1}_{a}$ denote the aggregated embedding of influence diffusion from the social neighbors
and $\tilde{\mathbf{q}}^{k+1}_{a}$ represents the embedding of aggregated interest diffusion from the interested item neighbors at the $(k+1)$-th layer. Then, each user's updated embedding $\mathbf{u}^{k+1}_a$ is modeled as:

\begin{small}
\vspace{-0.2cm}
\begin{flalign}
\label{eq:user_update}  & \mathbf{u}^{k+1}_a=\mathbf{u}^k_a+ (\gamma^{k+1}_{a1}\tilde{\mathbf{p}}^{k+1}_a+\gamma^{k+1}_{a2}\tilde{\mathbf{q}}^{k+1}_a), \\
\label{eq:user_socdiff} &\tilde{ \mathbf{p}}^{k+1}_{a} = \sum_{b\in S_a}\alpha^{k+1}_{ab}\mathbf{u}^k_{b}, \\
\label{eq:user_intdiff} &\tilde{\mathbf{q}}^{k+1}_{a} = \sum_{i\in R_a}\beta^{k+1}_{ai}\mathbf{v}^k_i,
\end{flalign}
\vspace{-0.4cm}
\end{small}

\noindent where Eq.\eqref{eq:user_update} shows how each user updates her latent embedding by fusing the influence diffusion aggregation $\tilde{\mathbf{p}}^{k+1}_a$  and interest diffusion aggregation $\tilde{\mathbf{q}}^{k+1}_a$, as well as her own embedding $\mathbf{u}^k_a$ at previous layer. Since each user appears in both the social graph and interest graph, Eq.\eqref{eq:user_socdiff} and Eq.\eqref{eq:user_intdiff} model the influence diffusion aggregation and interest diffusion aggregation from the two graphs respectively. Specifically, $\alpha^{k+1}_{ab}$ denotes the social influence of user $b$ to $a$ at the $(k+1)$-th layer in the social network, and $\beta^{k+1}_{ai}$ denotes the attraction of item $i$ to user $a$ at the $(k+1)$-th layer in the interest network.

In addition to the user and item embeddings, there are three groups of weights in the above three equations. A naive idea is to directly set equal values of each kind of weights, i.e., $\gamma^{(k+1)}_{a1}$=$\gamma^{(k+1)}_{a2}$=$\frac{1}{2}$, $\alpha^{(k+1)}_{ab}$=$\frac{1}{|S_a|}$, and $\beta^{(k+1)}_{ai}$=$\frac{1}{|R_a|}$. However, this simple idea could not well model the different kinds of weights in the user decision process. In fact, these three groups of weights naturally present a two-layer multi-level structure. Specifically, the social influence strengths and the interest strengths could be seen as node-level weights, which model how each user balances different neighboring nodes in each graph. By sending the aggregations of node level attention into Eq.\eqref{eq:user_update},  $\gamma^{k+1}_{al}$ is the graph level weight that learns to fuse and aggregate information from different graphs. Specifically, the graph layer weights are important as they model how each user balances the social influences and her historical records for user embedding. Different users vary, with some users are more likely to be swayed by the social network while the interests of others are quite stable. Therefore, the weights in the graph attention layer for eah user also need to be personally adapted.

As the three groups of weights represent a multi-level structure, we therefore use a multi-level attention network to model the attentive weights. Specifically, the graph attention network is designed to learn the contribution weight of each aspect when updating $a$'s embedding with different graphs, i.e., $\tilde{\mathbf{p}}^{k+1}_a$ and $\tilde{\mathbf{q}}^{k+1}_a$ in Eq.\eqref{eq:user_update}, and the node attention networks are designed to learn the attentive weights in each social graph and each interest graph respectively. Specifically, the social influence score $\alpha^{k+1}_{ab}$ is calculated as follows:

\begin{small}
\vspace{-0.2cm}
\begin{equation} \label{eq:user_agg_smlp}{}
\alpha^{k+1}_{ab} = MLP_2([\mathbf{u}^k_a, \mathbf{u}^k_b]).
\end{equation}
\vspace{-0.4cm}
\end{small}

In the above equation, the social influence strength $\alpha^{k+1}_{ab}$ takes the related two users' embeddings at the $k$-th layer as input, and sending these features into a MLP to learn the complex relationship between features for social influence strength learning. Without confusion, we omit the normalization step of all attention modeling in the following, as all of them share the similar form as shown in Eq.\eqref{eq:item_agg_normalize}.

Similarly, we calculate the interest influence score $\beta^{k+1}_{ai}$ by taking related user embedding and item embedding as input:

\begin{small}
\vspace{-0.2cm}
\begin{equation} \label{eq:user_agg_imlp}
\beta^{k+1}_{ai} = MLP_3([\mathbf{u}^k_a, \mathbf{v}^k_i]) \\
\end{equation}
\vspace{-0.4cm}
\end{small}

After obtaining the two groups of the node attentive weights, the output of the node attention weights are sent to the graph attention network, and we could model the graph attention weights of $\gamma^{k+1}_{al}(l=1, 2)$ as:

\begin{small}
\vspace{-0.2cm}
\begin{flalign}
\label{eq:user_agg_hsmlp}&\gamma^{k+1}_{a1} = MLP_4([\mathbf{u}^k_a, \tilde{\mathbf{p}}^k_a]) \\
\label{eq:user_agg_himlp}&\gamma^{k+1}_{a2} = MLP_4([\mathbf{u}^k_a, \tilde{\mathbf{q}}^k_a])
\end{flalign}
\vspace{-0.4cm}
\end{small}

In the above equation,  for each user $a$,  the graph attention layer scores not only rely on the user's embedding~($\mathbf{u}^k_a$), but also the weighted representations that are learnt from the node attention network. For example, as shown in Eq.\eqref{eq:user_update}, $\gamma^{(k+1)}_{a1}$ denotes the influence diffusion weight for contributing to users' depth $(k+1)$ embedding , with additional input of the
learned attentive combination of the influence diffusion aggregation in Eq.\eqref{eq:user_socdiff}. Similarly,  $\gamma^{(k+1)}_{a2}$ denotes the interest diffusion weight for contributing to users' depth $(k\!+\!1)$ embedding , with additional input of the learned attentive combination of the interest diffusion aggregation in Eq.\eqref{eq:user_intdiff}. As $\gamma^{k+1}_{a1}\!+\!\gamma^{k+1}_{a2}\!=\!1$, larger $\gamma^{k+1}_{a1}$ denotes higher influence diffusion effect with less interest diffusion effect. Therefore, the learned aspect importance scores are tailored to each user, which distinguish the importance of the influence diffusion effect and interest diffusion effect during the user's embedding updating process.

\textbf{Prediction Layer.} After the iterative $K$-layer diffusion process, we obtain the embedding set of $u$ and $i$ with $\mathbf{u}_{a}^{k}$ and $\mathbf{v}_{i}^{k}$ for $k=[0, 1, 2, ..., K]$. Then, for each user $a$, her final embedding is denoted as: $\mathbf{u}^*_a=[\mathbf{u}^0_a||\mathbf{u}^1_a||...||\mathbf{u}_{a}^{K}]$ that concatenates her embedding at each layer. Similarly, each item $i$'s final embedding is :  $\mathbf{v}^*_i=[\mathbf{v}^0_i||\mathbf{v}^1_i||...||\mathbf{v}_{i}^{K}]$. After that, the predicted rating is modeled as  the inner product between the final user and item embeddings~\cite{chen2020revisiting}:

\begin{equation} \label{output_prediction}
\begin{small}
       \hat{r}_{ai} = [\mathbf{u}^0_a||\mathbf{u}^1_a||...||\mathbf{u}_{a}^{K}]^T [\mathbf{v}^0_i||\mathbf{v}^1_i||...||\mathbf{v}_{i}^{K}].
\end{small}
\end{equation}

Please note that, some previous works directly use the $K$-th layer embedding for prediction layer as $\hat{r}_{ai}=[\mathbf{u}_a^K]^T\mathbf{V}_i^K$. Recently, researchers found that if we use the $K$-th layer embedding, GCN based approaches are proven to over-smoothing issue as $K$ increases~\cite{li2018deeper,zhou2018graph}. In this paper, to tackle the over-smoothing problem, we adopt the prediction layer as the LR-GCCF model, which receives state-of-the-art performance with user-item bipartite graph structure~\cite{chen2020revisiting}. In LR-GCCF, Chen et al. carefully analyzed the simple concatenation of entity embedding at each layer is equivalent to residual preference learning, and why this simple operation could alleviate the over-smoothing issue~\cite{chen2020revisiting}.

\subsection{Model Training}

We use a pair-wise ranking based loss function for optimization, which is widely used for implicit feedback~\cite{bpr}:

\vspace{-5pt}

\begin{equation} \label{eq:objective_function}
\begin{small}
L = \min_{\Theta} \sum_{(a, i) \in R^+ \cup (a,j)\in R^-}-ln\sigma(\hat{r}_{ai}-\hat{r}_{aj}) + \lambda||\Theta||^2~.
\end{small}
\end{equation}

\noindent where $R^+$ denotes the set of positive samples~(observed user-item pairs), and $R^-$ denotes the set of negative samples~(unobserved user-item pairs that randomly sampled from $\mathbf{\mathrm{R}}$). $\sigma(x)$ is sigmoid function. {\small$\Theta\!=\![\Theta_1, \Theta_2]$} is the regularization parameters in our model, with  {\small$\Theta_1\!=\![\mathbf{P},\mathbf{Q}]$}, and the parameter set in the fusion layer and the multi-level attention modeling, i.e.,  {\small $\Theta_2\!=\![\mathbf{W}_1, \mathbf{W}_2, [MLP_i]_{i=1,2,3,4}]$}. All the parameters in the above loss function are differentiable.

For all the trainable parameters, we initialize them with the Gaussian distribution with a mean value of 0 and a standard deviation of 0.01. Besides, we do not deliberately adjust the dimensions of each embedding size in the convolutional layer, all of them keep the same size. As for the several MLPs in the multi-level attention network, we use two-layer structure. In the experiment part, we will give more detail descriptions about the parameter setting.

\subsection{Matrix Formulation of DiffNet++}

The key idea of our proposed DiffNet++ model is the well designed interest and influence diffusion layers. In fact, this part could be calculated in matrix forms. In the following, we would like to show how to update user and item embedding from the $k$-th layer to the $(k+1)$-th layer with matrix operations. Let $\mathbf{H}^{(k+1)}=[\eta^{k+1}_{ia}]\in \mathbb{R}^{N\times M}$ denote the matrix representation of attentive item aggregation weigth in Eq.\eqref{eq:item_agg_mlp}, we have:

\begin{small}
\begin{equation} \label{eq:matrix_att}
\mathbf{H}=MLP_1(\mathbf{U}^k, \mathbf{V}^k).
\end{equation}
\end{small}

At the user side, given Eq.\eqref{eq:user_update} ,let $\mathbf{A}^{(k+1)}=[\alpha^{k+1}_{ab}]\in \mathbb{R}^{M\times M}$, $\mathbf{B}^{(k+1)}=[\beta^{k+1}_{ia}]\in \mathbb{R}^{M\times N}$ denote the attentive weight matrices of social network~(Eq.\eqref{eq:user_socdiff}) and interest network(Eq.\eqref{eq:user_intdiff}), i.e., the outputs of the node attention layer. We use $\mathbf{\Gamma}^{(k+1)}=[\gamma^{k+1}_{al}]\in \mathbb{R}^{M\times 2}$ to denote the attentive weight matrix of the multi-level networks in Eq.\eqref{eq:user_agg_hsmlp} and Eq.\eqref{eq:user_agg_himlp}. All these three attention matrices can be calculated similarly as shown above.

After learning the attention matrices, we could update user and item embeddings at the $(k+1)$-th layer as:

\begin{footnotesize}
\begin{flalign}
  &\begin{bmatrix}
   \mathbf{U}^{(k+1)}  \\
   \mathbf{V}^{(k+1)}  \\
  \end{bmatrix}
  =
  \begin{bmatrix}
     \mathbf{I}_1 & \mathbf{R}.*\mathbf{B}.*rm(\mathbf{\Gamma}(:,2),N) \\
     \mathbf{R}^T.*\mathbf{H} & \mathbf{I}_2 \\
  \end{bmatrix}
  \begin{bmatrix}
   \mathbf{U}^{(k)}  \\
   \mathbf{V}^{(k)}  \\
  \end{bmatrix}  \label{eq:matrix_diffi} \\
&\hspace{10mm}+
  \begin{bmatrix}
     \mathbf{S}.*\mathbf{A}.*rm(
     \mathbf{\Gamma(:,1)},M) & \mathbf{0} \\
     \mathbf{0}& \mathbf{0} \\
  \end{bmatrix}
  \begin{bmatrix}
   \mathbf{U}^{(k)}  \\
   \mathbf{V}^{(k)}  \\
  \end{bmatrix}  \label{eq:matrix_diffs} \\
  &=
  \begin{bmatrix}
     \mathbf{I}_1+\mathbf{S}.*\mathbf{A}.*rm(\mathbf{\Gamma}(:,1),M) & \mathbf{R}.*\mathbf{B}.*rm(\mathbf{\Gamma}(:,2),N) \\
     \mathbf{R}^T.*\mathbf{H} & \mathbf{I}_2\\
  \end{bmatrix}
  \begin{bmatrix}
   \mathbf{U}^{(k)}  \\
   \mathbf{V}^{(k)}  \\
  \end{bmatrix}, \label{eq:matrix_diffsi}
\end{flalign}
\end{footnotesize}

\noindent where {\small$\mathbf{I}_1$} is an identity matrix with $M$ rows, and {\small$\mathbf{I}_2$} is an identity matrix with $N$ rows. Moreover, $\mathbf{\Gamma}(:,1)$ and $\mathbf{\Gamma}(:,2)$ represent the first column and second column of matrix $\mathbf{\Gamma}$,  $.*$ denotes the dot product and $rm(\mathbf{A}, r_1)$ denotes an array that containing $r_1$ copies of  {\small$\mathbf{A}$} in the column dimensions.

Based on the above matrix operations of the social and influence diffusion layers, DiffNet++ is easily implemented by current deep learning frameworks.

\subsection{Discussion}

\textbf{Space complexity.} As shown in Eq.\eqref{eq:objective_function}, the model parameters are composed of two parts: the user and item free embeddings {\small$\Theta_1$=$[\mathbf{P},\mathbf{Q}]$}, and the parameter set in the fusion layer and the attention modeling, i.e.,  {\small $\Theta_2$=$[\mathbf{W}_1, \mathbf{W}_2, [MLP_i]_{i=1,2,3,4}]$}. Since most embedding based models~(e.g., BPR~\cite{bpr}, FM~\cite{rendle2010factorization}) need to store the embeddings of each user and each item, the space complexity of $\Theta_1$ is the same as classical embedding based models and grows linearly with users and items. For parameters in {\small$\Theta_2$}, they are shared among all users and items, with the dimension of each parameter is far less than the number of users and items. In practice, we empirically find the two-layer MLP achieve the best performance. As such, this additional storage cost is a small constant that could be neglected. Therefore, the space complexity of DiffNet++ is the same as classical embedding models.

\textbf{Time complexity.} Compared to the classical matrix factorization based models, the additional time cost lies in the influence and interest diffusion layers. Given $M$ users, $N$ items and diffusion depth $K$, suppose each user directly connects to $L_s$ users and $L_i$ items on average, and each item directly connects to $L_u$ users. At each influence and interest diffusion layer, we need to first calculate the two-level attention weight matrices as shown in Eq.\eqref{eq:matrix_att}, and then update user and item embeddings. Since in practice, MLP layers are very small~(e.g., two layers), the time cost for attention modeling is about $O(M(L_s+L_i)D+NL_uD)$. After that, as shown in Eq.\eqref{eq:matrix_diffsi}, the user and item update step also costs $O(M(L_s+L_i)D+NL_uD)$. Since there are $K$ diffusion layers, the total additional time complexity for influence and interest diffusion layers are $O(K(M(L_s+L_i)+NL_u)D)$. In practice, as $L_s, L_i, L_u\ll min\{M, N\}$, the additional time is linear with users and items, and grows linearly with diffusion depth $K$. Therefore, the total time complexity is acceptable in practice.

\textbf{Model Generalization.} The proposed DiffNet++ model is designed under the problem setting with the input of  user feature matrix {\small$\mathbf{X}$}, item feature matrix {\small$\mathbf{Y}$}, and the social network {\small$\mathbf{S}$}. Specifically, the fusion layer takes users'~(items') feature matrix   for user~(item) representation learning. The layer-wise diffusion layer utilizes the social network structure {\small$\mathbf{S}$} and the interest network structure {\small $\mathbf{R}$} to model how users' latent preferences are dynamically influenced from the recursive influence and interest diffusion process. Next, we would show that our proposed model is generally applicable when different kinds of data input are not available.

When the user~(item) features are not available, the fusion layer disappears. In other words, as shown in Eq.\eqref{eq:item_embed}, each item's latent embedding $\mathbf{v}^0_i$ degenerates to $\mathbf{q}_i$. Similarly, each user's initial layer-0 latent embedding  $\mathbf{u}^0\!=\!\mathbf{p}_a$ ~(Eq.\eqref{eq:fusion_user}).  Similarly, when either the user attributes or the item attributes do not exist, the corresponding fusion layer of user or item degenerates.

\begin{small}
\begin{table}[htbp!]\label{dataset}
\centering
\small
\caption{The statistics of the four datasets after preprocessing.}
\vspace{-3pt}
\begin{tabular}{|c|c|c|c|c|}
\hline
Dataset  & Yelp   &  Flickr    &  Epinions & Dianping\\ 
\hline

Users    & 17,237    &  8,358  &18,202  &59,426 \\
Items   & 38,342   &  82,120   &47,449  &10,224\\  \hline
Ratings & 204,448   &  327,815  &298,173 &934,334 \\
Links   & 143,765   &  187,273  &381,559 &813,331\\   \hline

Rating Density &0.03\% &0.05\%  & 0.03\% &0.12\% \\
Link Density  &0.05\% &0.27\%   & 0.15\%  &0.02\%\\
\hline
Attributes & $\surd$   & $\surd$         & $\times$    & $\times$      \\ \hline
\end{tabular}
\label{dataset}
\end{table}
\end{small}

\section{Experiments}

\textbf{Datasets.} We conduct experiments on four real-world datasets: \emph{Yelp}, \emph{Flickr}, \emph{Epinions} and \emph{Dianping}.

Yelp is a well-known online location based social network, where users could  make friends with others and review restaurants. We use the Yelp dataset that is publicly available\footnote[2]{https://www.yelp.com/dataset}. 
Flickr\footnote[3]{http://flickr.com/} is an online image based social sharing platform for users to follow others and share image preferences. In this paper, we use the  social image recommendation dataset that is crawled and published by authors in~\cite{wu2019hierarchical}, with both the social network structure and users' rating records of images. Epinions is a social based product review platform and the dataset is introduced in \cite{massa2007trust}  and is publicly available~\footnote[4]{http://www.trustlet.org/downloaded\_epinions.html}. Dianping is the largest Chinese location based social network, and we use this dataset that is crawled by authors in~\cite{li2015overlapping}. This dataset is also publicly available~\footnote[5]{https://lihui.info/data/}.

Among the four datasets, Yelp and Flickr are two datasets with user and item attributes, and are adopted as datasets of our previously proposed DiffNet model~\cite{leDiffnet}. The remaining two datasets of Epinions and Dianping do not contain user and item attributes. We use the same preprocessing steps of the four datasets. Specifically, as the original ratings are presented with detailed values, we transform the original scores to binary values. If the rating value is larger than 3, we transform it into 1, otherwise it equals 0. For both datasets, we filter out users that have less than 2 rating records and 2 social links and remove items which have been rated less than 2 times.
We randomly select 10\% of the data for the test. In the remaining 90\% data, to tune the parameters, we select 10\% from the training data as the validation set.  We show an overview of the characteristics of the four datasets in Table~\ref{dataset}. In this table, the last line shows whether the additional user and item attributes are available on this dataset.

\begin{table}[H]\label{character}
\centering
\setlength\tabcolsep{1.4pt}
\caption{Comparison of the baselines, with ``F" represents feature input and ``S" denotes the social network input.  For the modeling process, we use $OI$ and $OS$ to denote the observed first-order interest network and social network for user embedding learning. We use ``HS'' to denote the higher-order social information for embedding learning, and ``HI" to denote higher-order interest information for embedding learning. }
\begin{small}
\begin{tabular}{|l|l|p{1.0cm}|l|p{0.8cm}|p{0.8cm}|p{0.8cm}|l|}
\hline
\multicolumn{2}{|c|}{\multirow{2}{*}{Model}} & \multicolumn{2}{c|}{\footnotesize{Model Input}} & \multicolumn{4}{c|}{\footnotesize{User Embedding Ability}} \\ \cline{3-8}
\multicolumn{2}{|l|}{} & \multicolumn{1}{l|}{F} & S & OI & OS & HI & HS \\ \hline
\multirow{2}{*}{\begin{tabular}[c]{@{}l@{}}Classical \\ CF\end{tabular}} & BPR~\cite{bpr}& $\times$ & $\times$ & $\times$ & $\times$ & $\times$ & $\times$ \\ \cline{2-8}
 & FM~\cite{rendle2010factorization} & $\surd$ & $\times$ & $\times$ & $\times$ & $\times$ & $\times$ \\ \hline
\multirow{4}{*}{\begin{tabular}[c]{@{}l@{}}Social \\ recom-\\ mendation\end{tabular}} & SocialMF~\cite{socialmf} & $\times$ & $\surd$ & $\times$ & $\surd$ & $\times$ & $\times$ \\ \cline{2-8}
 & TrustSVD~\cite{trustsvd}  & $\times$ & $\surd$ & $\surd$ & $\surd$ & $\times$ & $\times$ \\ \cline{2-8}
 & ContextMF~\cite{tkde2014scalable}  & $\surd$ & $\surd$ & $\times$ & $\surd$ & $\times$ & $\times$ \\ \cline{2-8}
 & CNSR~\cite{wu2018collaborative} & $\times$ & $\surd$ & $\times$ & $\surd$ & $\times$ & $\surd$ \\ \hline
\multirow{7}{*}{\begin{tabular}[c]{@{}l@{}}Graph\\ neural\\ network\\ based\\recom-\\mendation \end{tabular}}  & GraphRec~\cite{fan2019graph} & $\times$ & $\surd$ & $\surd$ & $\surd$ & $\times$ & $\times$ \\ \cline{2-8}
 & PinSage~\cite{ying2018graph}  & $\surd$ & $\times$ & $\surd$ & $\times$ & $\surd$ & $\times$ \\ \cline{2-8}
 & NGCF~\cite{wang2019neural}  & $\times$ & $\times$ & $\surd$ & $\times$ & $\surd$ & $\times$ \\  \cline{2-8}
 & \emph{DiffNet-nf}~\cite{leDiffnet}  & $\times$ & $\surd$ & $\times$ & $\surd$ & $\times$ & $\surd$ \\  \cline{2-8}
 & \emph{DiffNet}~\cite{leDiffnet}    & $\surd$ & $\surd$ & $\times$ & $\surd$ & $\times$ & $\surd$ \\  \cline{2-8}
 & \emph{DiffNet++-nf } & $\times$ & $\surd$ & $\surd$ & $\surd$ & $\surd$ & $\surd$ \\  \cline{2-8}
 & \emph{DiffNet++ } & $\surd$ & $\surd$ & $\surd$ & $\surd$ & $\surd$ & $\surd$ \\ \hline
\end{tabular}
 \end{small}
\end{table}

\begin{table*}[htb!] \label{performance1}
\centering
\setlength{\tabcolsep}{1.5mm}
\caption{Overall comparison with different dimension size D on Yelp and Flickr (attributes are available).}
\vspace{-4pt}
\begin{small}
\begin{tabular}{|c|c|c|c|c|c|c|c|c|c|c|c|c|}
\hline
\multirow{3}{*}{Model} & \multicolumn{6}{c|}{Yelp}  & \multicolumn{6}{c|}{Flickr}   \\ \cline{2-13}
 & \multicolumn{3}{c|}{HR}   & \multicolumn{3}{c|}{NDCG} & \multicolumn{3}{c|}{HR}   & \multicolumn{3}{c|}{NDCG} \\ \cline{2-13}
 & \multicolumn{1}{c|}{D=16} & D=32 & D=64 & D=16  & D=32   & D=64  & \multicolumn{1}{c|}{D=16} & \multicolumn{1}{l|}{D=32} & \multicolumn{1}{l|}{D=64} & D=16    & D=32   & D=64   \\ \hline

BPR	        &0.2435&0.2616&0.2632&0.1468&0.1573&0.1554&0.0773&0.0812&0.0795&0.0611&0.0652&0.0628\\ \hline
FM       &0.2768&0.2835&0.2825&0.1698&0.1720&0.1717&0.1115&0.1212&0.1233&0.0872&0.0968&0.0954	\\ \hline \hline
SocialMF	&0.2571&0.2709&0.2785&0.1655&0.1695&0.1677&0.1001&0.1056&0.1174&0.0862&0.0910&0.0964\\ \hline
TrustSVD    &0.2826&0.2854&0.2939&0.1683&0.1710&0.1749&0.1352&0.1341&0.1404&0.1056&0.1039&0.1083\\ \hline
ContextMF   &0.2985&0.3011&0.3043&0.1758&0.1808&0.1818&0.1405&0.1382&0.1433&0.1085&0.1079&0.1102\\ \hline
CNSR       &0.2702&0.2817&0.2904&0.1723&0.1745&0.1746&0.1146&0.1198&0.1229&0.0913&0.0942&0.0978	\\ \hline \hline
GraphRec    &0.2873&0.2910&0.2912&0.1663&0.1677&0.1812&0.1195&0.1211&0.1231&0.0910&0.0924&0.0930\\ \hline
PinSage     &0.2944&0.2966&0.3049&0.1753&0.1786&0.1855&0.1192&0.1234&0.1257&0.0937&0.0986&0.0998\\ \hline
NGCF        &0.3050&0.3068&0.3042&0.1826&0.1844&0.1828&0.1110&0.1150&0.1189&0.0880&0.0895&0.0945\\ \hline \hline
DiffNet-nf  &0.3126&0.3156&0.3195&0.1854&0.1882&0.1928&0.1342&0.1317&0.1408&0.1040&0.1034&0.1089\\ \hline
DiffNet     &0.3293&0.3437&0.3461&0.1982&0.2095&0.2118&0.1476&0.1588&0.1657&0.1121&0.1242&0.1271\\ \hline
{DiffNet++-nf}    &0.3194&0.3199&0.3230&0.1914&0.1944&0.1942&0.1410&0.1480&0.1503&0.1100&0.1132&0.1169\\ \hline

DiffNet++      &\textbf{0.3406}&\textbf{0.3552}&\textbf{0.3694}&\textbf{0.2070}&\textbf{0.2158}&\textbf{0.2263}&\textbf{0.1562}&\textbf{0.1678}&\textbf{0.1832}&\textbf{0.1213}&\textbf{0.1286}&\textbf{0.1420}\\ \hline
\end{tabular}
\end{small}
\label{performance1}
\end{table*}

\begin{table*}[htb!] \label{performance1_new}
\centering
\setlength{\tabcolsep}{1.5mm}
\caption{Overall comparison with different dimension size D on Epinions and Dianping (attributes are not available).}
\vspace{-4pt}
\begin{small}
\begin{tabular}{|c|c|c|c|c|c|c|c|c|c|c|c|c|}
\hline
\multirow{3}{*}{Model} & \multicolumn{6}{c|}{Epinions}  & \multicolumn{6}{c|}{Dianping}   \\ \cline{2-13}
 & \multicolumn{3}{c|}{HR}   & \multicolumn{3}{c|}{NDCG} & \multicolumn{3}{c|}{HR}   & \multicolumn{3}{c|}{NDCG} \\ \cline{2-13}
 & \multicolumn{1}{c|}{D=16} & D=32 & D=64 & D=16  & D=32   & D=64  & \multicolumn{1}{c|}{D=16} & \multicolumn{1}{l|}{D=32} & \multicolumn{1}{l|}{D=64} & D=16    & D=32   & D=64   \\ \hline

BPR&	       0.2620& 0.2732& 0.2822 &0.1702 &0.1788 &0.1812  & 0.2160 &  0.2302&	0.2299&	0.1286&	0.1326& 0.1319
 \\ \hline \hline
SocialMF&   0.2720 & 0.2842& 0.2893 &0.1732 &0.1824 &0.1857 & 0.2325& 0.2345&	0.2410&	0.1360&	0.1377&	0.1416
 \\ \hline
TrustSVD&   0.2726 &0.2854 &0.2884 &0.1773& 0.1839& 0.1848  & 0.2364& 0.2371&	0.2341&	0.1381&	0.1401&	0.1390
  \\ \hline
CNSR&       0.2757 &0.2874 &0.2898 & 0.1748& 0.1856& 0.1876 & 0.2356& 0.2377&	0.2418&	0.1394&	0.1413&	0.1435
 \\ \hline \hline
GraphRec&   0.3093 &0.3117 &0.3156 &0.1994 &0.2016 &0.2051  & 0.2408& 0.2541&	0.2622&	0.1412&	0.1503&	0.1556
  \\ \hline
PinSage&    0.2980 &0.3003 &0.3073 &0.1911 &0.1933 &0.1928  & 0.2353& 0.2452&	0.2552&	0.1390&	0.1434&	0.1489
  \\ \hline
NGCF&       0.3029 &0.3065 &0.3192 &0.1977 &0.2008 &0.1958  & 0.2489& 0.2586&	0.2584&	0.1470&	0.1503&	0.1534
  \\ \hline \hline
DiffNet&    0.3242 &0.3281 &0.3407 &0.2007 &0.2054 &0.2191   & 0.2522& 0.2600&	0.2645&	0.1483&	0.1521&	0.1555
 \\ \hline
{DiffNet++}&
  \textbf{0.3367}	&\textbf{0.3434}	&\textbf{0.3503}	&\textbf{0.2158}	&\textbf{0.2217}	&\textbf{0.2288} & \textbf{0.2676}& \textbf{0.2682}&	\textbf{0.2713}&	\textbf{0.1593}&	\textbf{0.1589}&	\textbf{0.1605}
 \\ \hline
\end{tabular}
\end{small}
\label{performance1_new}
\end{table*}

\begin{table*}[htb!]\label{performance2}
\centering
\setlength{\tabcolsep}{1.5mm}
\caption{ Overall comparison with different top-N values~(D=64) on Yelp and Flickr (attributes are available). }
\vspace{-4pt}
\begin{small}
\begin{tabular}{|c|c|c|c|c|c|c|c|c|c|c|c|c|}
\hline
\multirow{3}{*}{Model} & \multicolumn{6}{c|}{Yelp}  & \multicolumn{6}{c|}{Flickr}   \\ \cline{2-13}
 & \multicolumn{3}{c|}{HR}   & \multicolumn{3}{c|}{NDCG} & \multicolumn{3}{c|}{HR}   & \multicolumn{3}{c|}{NDCG} \\ \cline{2-13}
 & \multicolumn{1}{c|}{N=5} & N=10 & N=15 & N=5  & N=10   & N=15  & \multicolumn{1}{c|}{N=5} & \multicolumn{1}{c|}{N=10} & \multicolumn{1}{c|}{N=15} & N=5  & N=10   & N=15    \\ \hline
BPR	        &0.1695&0.2632&0.3252&0.1231&0.1554&0.1758&0.0651&0.0795&0.1037&0.0603&0.0628&0.0732\\ \hline
FM	        &0.1855&0.2825&0.3440&0.1341&0.1717&0.1876&0.0989&0.1233&0.1473&0.0866&0.0954&0.1062\\ \hline\hline
SocialMF    &0.1739&0.2785&0.3365&0.1324&0.1677&0.1841&0.0813&0.1174&0.1300&0.0723&0.0964&0.1061\\ \hline
TrustSVD	&0.1882&0.2939&0.3688&0.1368&0.1749&0.1981&0.1089&0.1404&0.1738&0.0978&0.1083&0.1203\\ \hline
ContextMF	&0.2045&0.3043&0.3832&0.1484&0.1818&0.2081&0.1095&0.1433&0.1768&0.0920&0.1102&0.1131\\ \hline
CNSR	    &0.1877&0.2904&0.3458&0.1389&0.1746&0.1912&0.0920&0.1229&0.1445&0.0791&0.0978&0.1057\\ \hline\hline
GraphRec    &0.1915&0.2912&0.3623&0.1279&0.1812&0.1956&0.0931&0.1231&0.1482&0.0784&0.0930&0.0992\\ \hline
PinSage     &0.2105&0.3049&0.3863&0.1539&0.1855&0.2137&0.0934&0.1257&0.1502&0.0844&0.0998&0.1046\\ \hline
NGCF        &0.1992&0.3042&0.3753&0.1450&0.1828&0.2041&0.0891&0.1189&0.1399&0.0819&0.0945&0.0998\\ \hline \hline
DiffNet-nf     &0.2101&0.3195&0.3982&0.1535&0.1928&0.2164&0.1087&0.1408&0.1709&0.0979&0.1089&0.1192\\ \hline
DiffNet     &0.2276&0.3461&0.4217&0.1679&0.2118&0.2307&0.1178&0.1657&0.1855&0.1072&0.1271&0.1301\\ \hline
DiffNet++-nf     &0.2112&0.3230&0.3989&0.1551&0.1942&0.2176&0.1140&0.1503&0.1799&0.1021&0.1169&0.1256\\ \hline
DiffNet++      &\textbf{0.2503}&\textbf{0.3694}&\textbf{0.4493}&\textbf{0.1841}&\textbf{0.2263}&\textbf{0.2497}&\textbf{0.1412}&\textbf{0.1832}&\textbf{0.2203}&\textbf{0.1269}&\textbf{0.1420}&\textbf{0.1544}\\ \hline

\end{tabular}
\end{small}
\label{performance2}
\end{table*}

\begin{table*}[htb!]\label{performance2_new}
\centering
\setlength{\tabcolsep}{1.5mm}
\caption{Overall comparison with different top-N values~(D=64)on Epinions and Dianping (attributes is not available).}
\vspace{-4pt}
\begin{small}
\begin{tabular}{|c|c|c|c|c|c|c|c|c|c|c|c|c|}
\hline
\multirow{3}{*}{Model} & \multicolumn{6}{c|}{Epinions}  & \multicolumn{6}{c|}{Dianping}   \\ \cline{2-13}
 & \multicolumn{3}{c|}{HR}   & \multicolumn{3}{c|}{NDCG} & \multicolumn{3}{c|}{HR}   & \multicolumn{3}{c|}{NDCG} \\ \cline{2-13}
 & \multicolumn{1}{c|}{N=5} & N=10 & N=15 & N=5  & N=10   & N=15  & \multicolumn{1}{c|}{N=5} & \multicolumn{1}{c|}{N=10} & \multicolumn{1}{c|}{N=15} & N=5  & N=10   & N=15    \\ \hline

BPR	 &     0.2005& 0.2822& 0.3256& 0.1526& 0.1812& 0.1917&  0.1412&	0.2299&	0.2864&	0.1024&	0.1319&	0.1482
 \\ \hline \hline
SocialMF&  0.2098& 0.2893& 0.3431& 0.1575& 0.1857& 0.2016&  0.1546&	0.2410&	0.3063&	0.1111&	0.1416&	0.1608
 \\ \hline
TrustSVD&  0.2102& 0.2884& 0.3396& 0.1574& 0.1848& 0.2001&  0.1521&	0.2341&	0.2966&	0.1100&	0.1390&	0.1574
  \\ \hline
CNSR&      0.2151& 0.2898& 0.3444& 0.1592& 0.1876& 0.2035&  0.1564&  0.2418&	0.3077&	0.1132&	0.1435&	0.1621
 \\ \hline \hline
GraphRec&  0.2335& 0.3156& 0.3620& 0.1764& 0.2051& 0.2199&  0.1725&	0.2622&	0.3300&	0.1240&	0.1556&	0.1755
  \\ \hline
PinSage&   0.2207& 0.3073& 0.3073& 0.1589& 0.1908& 0.2008&  0.1631&	0.2552&	0.3177&	0.1141&	0.1489&	0.1664
  \\ \hline
NGCF&      0.2308& 0.3192& 0.3777& 0.1706& 0.1958& 0.2131&  0.1695&	0.2584&	0.3263&	0.1220&	0.1534&	0.1733
  \\ \hline \hline
DiffNet&   0.2457& 0.3407& 0.3967& 0.1857& 0.2191& 0.2357&  0.1734&	0.2645&	0.3302&	0.1235&	0.1555&	0.1748
 \\ \hline
{DiffNet++}& \textbf{0.2602}&	\textbf{0.3503}&	\textbf{0.4051}&	\textbf{0.1973}&	\textbf{0.2288}&	\textbf{0.2450}& \textbf{0.1798}&	\textbf{0.2713}&	\textbf{0.3375}&	\textbf{0.1281}&	\textbf{0.1605}&	\textbf{0.1802}
 \\ \hline

\end{tabular}
\end{small}
\label{performance2_new}
\end{table*}

\textbf{Baselines and Evaluation Metrics.} To illustrate the effectiveness of our method, we compare DiffNet++ with competitive baselines, including classical CF models~(BPR~\cite{bpr}, FM~\cite{rendle2010factorization}), social based recommendation model~(SocialMF~\cite{socialmf}, TrustSVD~\cite{trustsvd}, ContextMF~\cite{tkde2014scalable}, CNSR~\cite{wu2018collaborative}), as well as the graph based recommendation models of GraphRec~\cite{fan2019graph}, PinSage~\cite{ying2018graph}, NGCF~\cite{wang2019neural}. Please note that, in PinSage, we take the user-item graph with both user and item features as input, in order to transform this model for the recommendation task. For our proposed models of DiffNet~\cite{leDiffnet} and DiffNet++, since both models are flexible and could be reduced to simpler versions without user and item features,  we use \mbox{DiffNet-nf} and \mbox{DiffNet++-nf} to represent reduced versions of DiffNet and DiffNet++ when removing user and item features.  For better illustration, we list the main characteristics of all these models in Table~\ref{character}, with our proposed models are listed with italic letters. Please note that, as BPR learns free user and item embeddings with the observed user-item ratings. Therefore, the first-order interest network is not learned in the emebedding modeling process. As can be seen from this paper, our proposed DiffNet++-nf and DiffNet++ are the only two models that consider both the higher-order social influence and higher-order interest network for social recommendation.

For the top-N ranking evaluation, we use two widely used metrics, Hit Ratio~(HR)~\cite{deshpande2004item} and Normalized Discounted Cummulative Gain~(NDCG)~\cite{deshpande2004item,leDiffnet}. Specifically,
HR measures the percentage of hit items in the top-N list, and NDCG puts more emphasis on the top ranked items.
As we focus on the top-N ranking performance with large itemset, similar as many other works~\cite{nmf,leDiffnet}, to evaluate the performance, for each user, we randomly select 1000 unrated items that a user has not interacted with as negative samples. Then, we mix these pseudo negative samples and corresponding positive samples~(in the test set) to select top-N potential candidates. To reduce the uncertainty in this process, we repeat this procedure 5 times and report the average results.

\textbf{Parameter Setting.} For the regularization parameter $\lambda$ in Eq.\eqref{eq:objective_function}, we empirically try it in the range of [0.0001, 0.001, 0.01, 0.1] and finally set $\lambda\!=\!0.01$  to get the best performance. For the fusion layer in Eq.\eqref{eq:fusion_user} and Eq.\eqref{eq:item_embed}, we first transform the each user~~(item) feature vector to the same free embedding space, and calculate as: $\mathbf{u}^0_a=\mathbf{W}_1\times\mathbf{x}_a+\mathbf{p}_a$, and $\mathbf{v}^0_a=\mathbf{W}_2\times\mathbf{y}_i+\mathbf{q}_a$. For attention modeling, we resort to MLP with two layers. For our proposed model, we initialize all of them with a Gaussian distribution with a mean value of 0 and the standard deviation of 0.01. We use the Adam optimizer for with an initial learning rate of 0.001, and the training batch size is 512. In the training process, as there are much more unobserved items for each user, we randomly select 8 times pseudo negative samples for each user at each iteration. Since each iteration we change the pseudo negative samples, each unobserved item gives very weak signal. For all the baselines, we carefully tune the parameters to ensure the best performance.

\begin{figure*}[htbp!]
\centering
\subfigure[Yelp dataset.]{
\centering
\includegraphics[width=75mm]{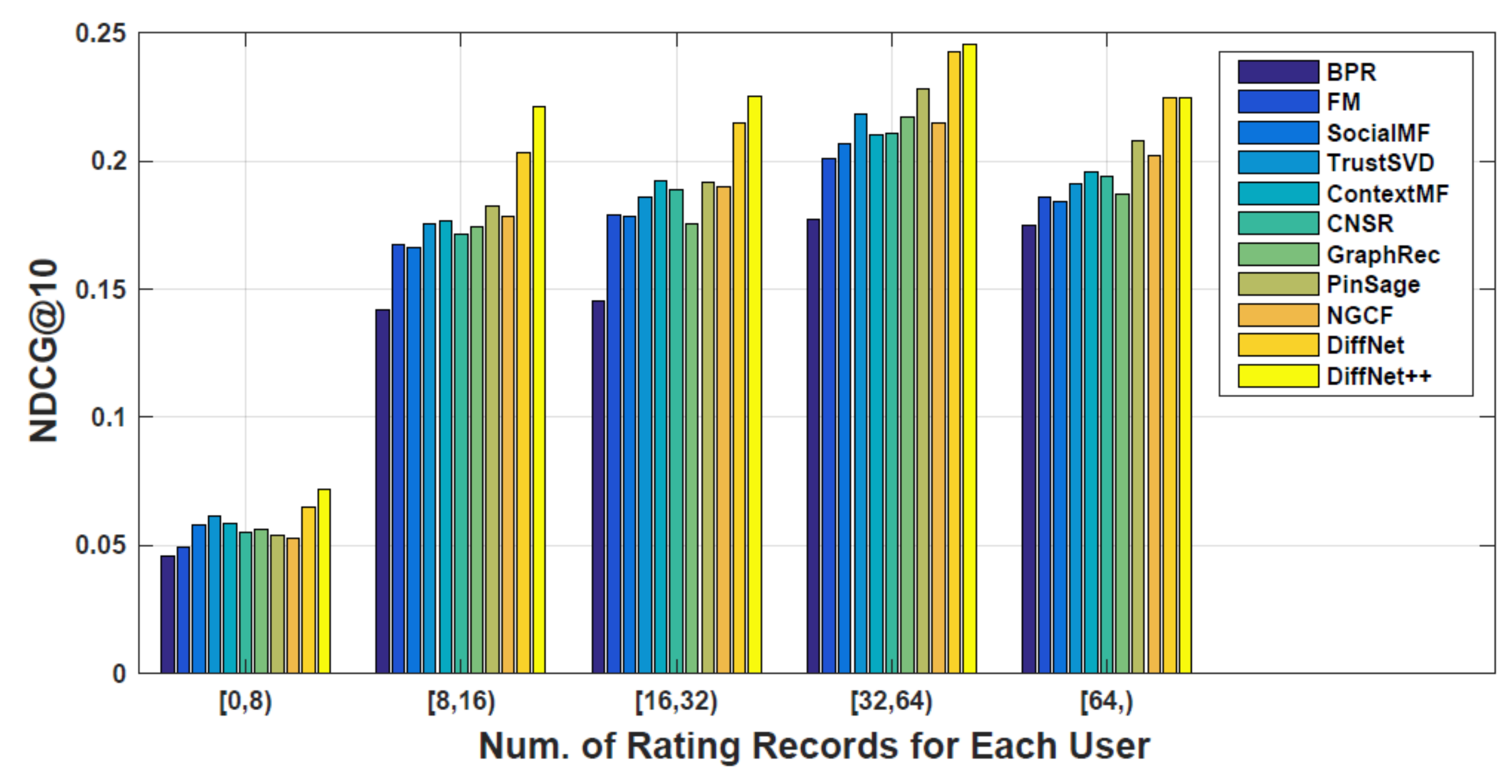}\label{fig:yelp_rating_sparsity}}
\subfigure[Flickr dataset.]{
\includegraphics[width=75mm]{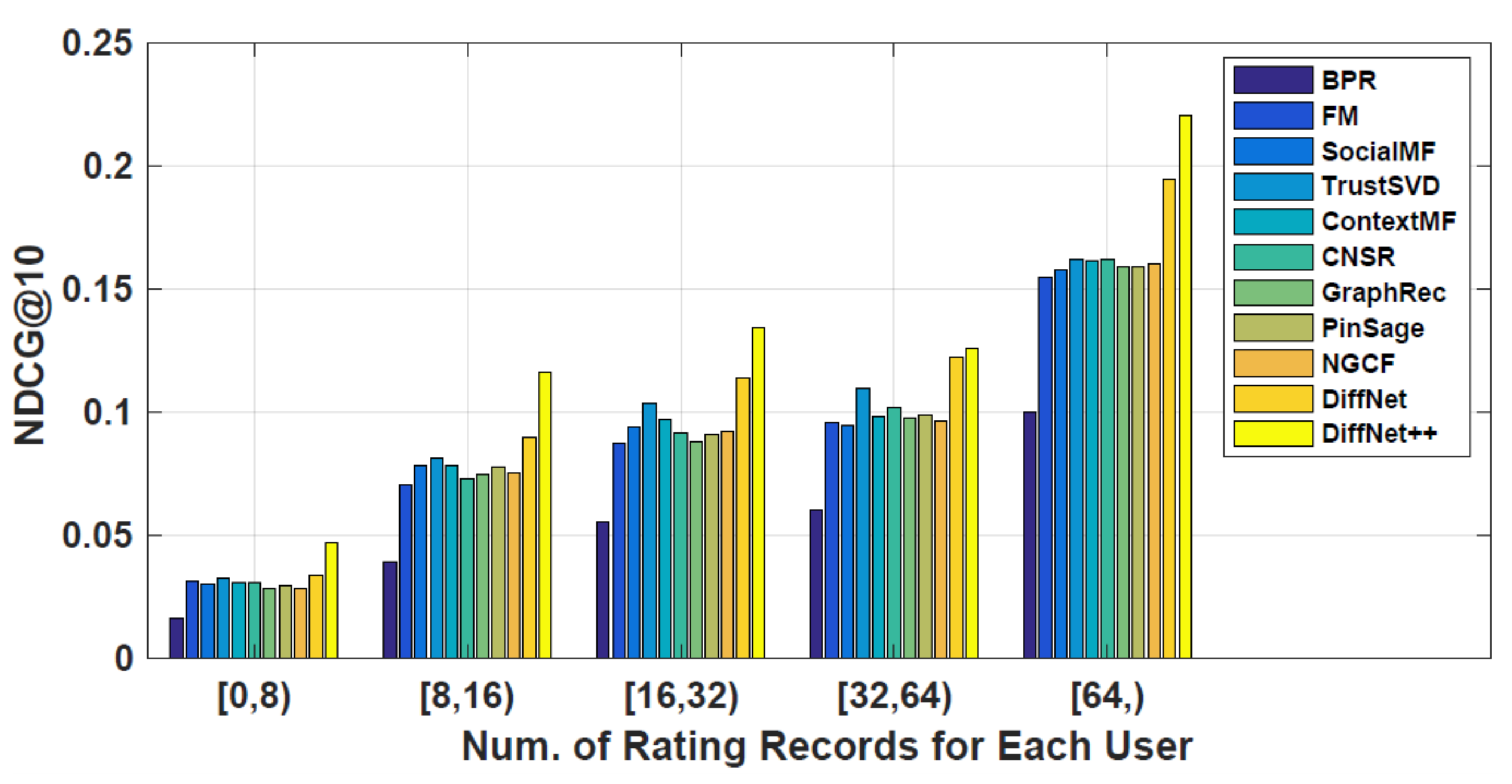}\label{fig:flickr_rating_sparsity}}
\centering
\caption{\footnotesize Performance under different rating sparsity on two datasets.} \label{fig:sparsity}
\end{figure*}

\subsection{Overall Performance Comparison}
We show the overall performance of all models for top-10 recommendation with different embedding size $D$ from Table~\ref{performance1} to Table\ref{performance2_new}. In Table~\ref{performance1}, we show the comparisons on Yelp and Flickr, with the node attribute values are available. In Table~\ref{performance1_new}, we depict the results on Epinions and Dianping without attribute values. On Epinions and Dianping, we do not report models that need to take attribute data as input. We notice that besides BPR, nearly all models show better performance with the increase of dimension D. All models improve over BPR, which only leverages the observed user-item rating matrix for recommendation, and suffer the data sparsity issue in practice. TrustSVD and SocialMF utilize  social neighbors of each user as  auxiliary information to alleviate this problem. GraphRec further improves over these traditional social recommendation models by jointly considering the first-order social neighbors and interest neighbors in the user embedding process. However, GraphRec only models the first-order relationships of two graphs for user embedding learning, with the higher-order graph structures are neglected. For GCN based models, PinSage and NGCF model the higher-order user-item graph structure, and DiffNet models the higher-order social structure. These graph neural models beat matrix based baselines by a large margin, showing the effectiveness in leveraging higher-order graph structure for recommendation. Our proposed DiffNet++ model always performs the best under any dimension $D$, indicating the effectiveness of modeling the recursive diffusion process in the social interest network. Besides, we observe DiffNet++ and DiffNet always show better performance compared to their counterparts that do not model the user and item features, showing the effectiveness of injecting both feature and latent embeddings in the fusion layer. We further compare the performance of different models with different top-N values in Table~\ref{performance2} and Table~\ref{performance2_new}, and the overall trend is the same as analyzed before. Therefore, we could empirically conclude the superiority of our proposed models. As nearly all models showed better performance at {\small$D\!=\!64$}, we would use this setting to show the comparison of different models in the following analysis.

\begin{table*}[htbp!] \label{performance_layer1}
\centering
\caption{HR@10 and NDCG@10 performance with different diffusion depth K~($D\!=\!64$).}
\vspace{-4pt}
\setlength{\tabcolsep}{2mm}
\begin{small}
\begin{tabular}{|c|c|c|c|c|c|c|c|c|}
\hline
\multirow{2}{*}{Depth $K$} & \multicolumn{4}{c|}{Yelp}   & \multicolumn{4}{c|}{Flickr}     \\ \cline{2-9}
 & \multicolumn{1}{c|}{HR} & Improve & \multicolumn{1}{c|}{NDCG} & Improve & \multicolumn{1}{c|}{HR} & Improve & \multicolumn{1}{c|}{NDCG} & Improve \\ \hline
$K$ = 2  &0.3694    &-          &0.2263    &-          &0.1832    &-            &0.1420    & -         \\ \hline
$K$ = 0  &0.2632    &-28.32\%   &0.1554    &-30.81\%   &0.0795    &-55.21\%     &0.0628    & -53.86\%   \\ \hline
$K$ = 1  &0.3566    &-2.89\%    &0.2159    &-3.87\%    &0.1676    &-5.58\%      &0.1283    & -5.73\%   \\ \hline
$K$ = 3  &0.3626    &-1.25\%    &0.2215    &-1.38\%    &0.1743    &-1.80\%      &0.1347    & -1.03\%   \\ \hline
\end{tabular}
\end{small}
\label{performance_layer1}
\end{table*}

\subsection{Performance Under Different Sparsity}

In this part, we would like to investigate how different models perform under different rating sparsity. Specifically, we first group users into different interest groups based on the number of observed ratings of each user. E.g, $[8, 16)$ means each user has at least 8 rating records and less than 16 rating records. Then, we calculate the average performance of each interest group. The sparsity analysis on Yelp dataset and Flickr dataset are  shown in Fig.~\ref{fig:yelp_rating_sparsity} and Fig.~\ref{fig:flickr_rating_sparsity} respectively. From both datasets, we observe that as users have more ratings, the  overall performance increases among all models. This is quite reasonable as all models could have more user behavior data for user embedding modeling.  Our proposed models consistently improve all baselines, and especially show larger improvements on sparser dataset. E.g., when users have less than 8 rating records, DiffNet++ improves 22.4\%  and 45.0\% over the best baseline on Yelp and Flickr respectively.

\subsection{Detailed Model Analysis}

\textbf{Diffusion Depth $K$}. The number of layer $K$ is very important, as it determines the diffusion depth of different graphs. We show the results of different $K$ values for two datasets with attributes in Table~\ref{performance_layer1}. The column of ``Improve'' shows the performance changes compared to the best setting, i.e., $K\!=\!2$. When $K$ increases from 0 to 1, the performance increases quickly~(DiffNet++ degenerates to BPR when $K\!=\!0$), and it achieves the best performance when $K\!=\!2$. However, when we continue to increase the layer to 3, the performance drops. We empirically conclude that 2-hop higher-order social interest graph structure is enough for social recommendation. And adding more layers may introduce unnecessary neighbors in this process, leading to performance decrease. Other related studies have also empirically found similar trends~\cite{kipf2016semi,ying2018graph}. 

\begin{table*}[htbp!]\label{tab:performance_att1}
\centering
\caption{HR@10 and NDCG@10 performance with different attentional variants~($D$ = 64).}
\vspace{-4pt}
\setlength{\tabcolsep}{1.5mm}
\begin{small}
\begin{tabular}{|c|c|c|c|c|c|c|c|c|c|}
\hline
\multirow{2}{*}{\begin{tabular}[c]{@{}c@{}}Graph Attention\end{tabular}} & \multirow{2}{*}{\begin{tabular}[c]{@{}c@{}}Node Attention\end{tabular}} & \multicolumn{4}{c|}{Yelp}   & \multicolumn{4}{c|}{Flickr}     \\ \cline{3-10}
  &   & \multicolumn{1}{c|}{HR} & \multicolumn{1}{c|}{Improve} & \multicolumn{1}{c|}{NDCG} & \multicolumn{1}{c|}{Improve}& \multicolumn{1}{c|}{HR} & \multicolumn{1}{c|}{Improve} & \multicolumn{1}{c|}{NDCG} & \multicolumn{1}{c|}{Improve} \\ \hline
AVG & AVG  &0.3631& - 		  &0.2224& - 	  	    &0.1733& - 	     &0.1329&-    		\\ \hline
AVG & ATT  &0.3657& +0.72\%   &0.2235& +0.49\%  	&0.1792& +3.40\% &0.1368&+2.93\%   \\ \hline
ATT & AVG  &0.3662& +0.85\%   &0.2249& +1.12\%  	&0.1814& +4.67\% &0.1387&+4.36\%	\\ \hline
ATT & ATT  &0.3694& +1.74\%   &0.2263& +1.75\%   	&0.1832& +5.71\% &0.1420&+6.85\%	\\ \hline

\end{tabular}
\end{small}
\label{performance_att1}
\end{table*}

\begin{table}[htbp!] \label{tab:att_stat}
\centering
\caption{Mean statistics of the graph level attention values~($K\!=\!2$), with $\gamma^{k}_{a1}$ is the social influence weight and $\gamma^{k}_{a2}$ is the interest weight.}
\vspace{-4pt}
\setlength{\tabcolsep}{2mm}
\begin{small}
\begin{tabular}{|c|c|c|c|c|}
\hline
\multirow{2}{*}{Layer $k$} & \multicolumn{2}{c|}{Yelp}   & \multicolumn{2}{c|}{Flickr}     \\ \cline{2-5}
 & Social $\gamma^k_{a1}$   & Interest  $\gamma^k_{a2}$
 & Social $\gamma^k_{a1}$ & Interest $\gamma^k_{a1}$
  \\ \cline{1-5}
$k$=1  &0.7309        &0.2691         &0.8381          &0.1619      \\ \hline
$k$=2  &0.6888        &0.3112         &0.0727          &0.9273        \\ \hline
\end{tabular}
\end{small}
\label{tab:att_stat}
\end{table}

\textbf{The Effects of Multi-level Attention.}
A key characteristic of our proposed model is the multi-level attention modeling by fusing the social network and interest network for recommendation. In this subsection, we discuss the effects of different attention mechanisms.
We show the results of different attention modeling combinations in in Table~\ref{performance_att1}, with ``AVG'' means we directly set the equal attention weights without any attention learning process. As can be observed from this table, either the node level attention or the graph level attention modeling could improve the recommendation results, with the graph level attention shows better results. When combing both the node level attention and the graph level attention, the performance can be further improved. E.g., for the Flickr dataset, the graph level attention improves more than 4\% compared to the results of the average attention, and combining the node level attention further improves about 2\%. However, the improvement of attention modeling varies in different datasets, with the results of the Yelp dataset is not as significant as the Flickr dataset. This observation implies that the usefulness of considering the importance strength of different elements in the modeling process varies, and our proposed multi-level attention modeling could adapt to different datasets' requirements.

\textbf{Attention Value Analysis.} For each user $a$ at layer $k$, the graph level attention weights of $\gamma^k_{a1}$ and $\gamma^k_{a2}$ denote the social influence diffusion weight and the interest diffusion weight. A larger value of $\gamma^k_{a1}$ indicates the social influence diffusion process is more important to capture the user embedding learning with less influence from the interest network. In Table~\ref{tab:att_stat},  we show the learned mean and variance of all users' attention weights at the graph level at each layer $k$. Since both datasets receive the best performance at {\small$K\!=\!2$}, we show the attention weights at the first diffusion layer~($k=1$) and the second diffusion layer $k\!=\!2$. There are several interesting findings. First, we observe for both datasets, at the first diffusion layer with $k\!=\!1$, the average value of the social influence strength $\gamma^1_{a1}$ are very high, indicating the first-order social neighbors play a very important role in representing each user's first layer representation. This is quite reasonable as users' rating behavior are very sparse, and leveraging the first order social neighbors could largely improve the recommendation performance. When $k\!=\!2$, the average social influence strength $\gamma^2_{a1}$ varies among the two datasets, with the Yelp dataset shows a larger average social influence weight, while the Flickr dataset shows a larger interest influence weight with quite small value of average social influence weight. We guess a possible reason is that, as shown in Table~\ref{dataset}, Flickr dataset shows denser social links compared to the Yelp dataset, with a considerable amount of directed social links at the first diffusion layer, the average weight of the second layer social neighbors decreases.

\textbf{Runtime.} In Table~\ref{tab:runtime}, we show the runtime of each model on four datasets. Among the four datastes, Epinions and Dianping do not have any attribute information.  For fair comparison, we perform experiments on a same server. The server has an Intel i9 CPU, 2 Titan RTX 24G, and 64G memory. The classical BPR model costs the least time, followed by the shallow latent factor based social recommendation models of SocailMF and TrustSVD. CNSR has longer runtime as it needs to update both the user embedding learned from user-item behavior, as well as the social embedding learned from user-user behavior. The graph based models cost more time than classical models. Specifically, NGCF and DiffNet have similar time complexity as they capture either the interest diffusion or influence diffusion. By injecting both the interest diffusion and influence diffusion process, DiffNet++ costs more time than these two neural graph models. GraphRec costs the most time on the two datasets without attributes. The reason is that, though GraphRec only considers one-hop graph structure, it adopts a deep neural architecture for modeling the complex interactions between users and items. As we need to use the deep neural architecture for each user-item rating record, GraphRec costs more time than the inner-product based prediction function in DiffNet++. On Yelp and Flickr, these two datasets have attribute information as input, and the DiffNet++ model needs the fusion layer to fuse attribute and free embeddings, while GraphRec does not have any attribute fusion. Therefore, DiffNet++ costs more time than GraphRec on the two datasets with attributes. The average training time of DiffNet++ on the largest Dianping dataset is about 25 seconds for one epoch, and it usually takes less than 100 epoches to reach convergence. Therefore, the total runtime of DiffNet++ is less than 1 hour on the largest dataset, which is also very time efficient.

\begin{table}[htbp!] \label{tab:runtime}
\centering
\caption{Average one epoch runtime of each model on the two largest datasets~(seconds).}
\vspace{-4pt}
\setlength{\tabcolsep}{2mm}
\begin{footnotesize}
\begin{tabular}{|c|c|c|c|c|}\hline
Model        &   Yelp & Flickr  &      Epinions & Dianping \\ \hline
BPR          &   1.51 & 1.75 & 2.34    &3.86 \\ \hline
FM           &   1.94 & 1.87 & $ \backslash $   & $ \backslash $   \\ \hline \hline
SocailMF     &   1.74 &3.38  &  2.41   &6.83 \\ \hline
TrustSVD     &   1.90 &3.44  &  2.60    &8.10  \\ \hline
ContextMF    &   1.83 & 3.51 & $ \backslash $   & $ \backslash $   \\ \hline
CNSR         &   2.55 &3.52  &  4.26    &13.24 \\ \hline \hline
GraphRec     &   4.33 & 5.65 &5.98    &42.98 \\ \hline
PinSage      &   4.07 & 3.58 &3.48    &19.28 \\ \hline
NGCF         &   4.07 &3.59&    3.28    &20.38 \\ \hline \hline
DiffNet      &   2.65 &3.42 &  3.15    &15.66  \\ \hline
DiffNet++    &   7.72 & 7.21 &  4.69    &25.62  \\ \hline
\end{tabular}
\end{footnotesize}
\label{tab:runtime}
\end{table}

\section{Conclusions and Future Work}
In this paper, we presented a neural social and interest diffusion based  model, i.e., DiffNet++, for social recommendation.
We argued that, as users play a central role in social network and interest network, jointly modeling the higher-order structure of these two networks would mutually enhance each other. By formulating the social recommendation as a heterogeneous graph, we recursively learned the user embedding from convolutions on user social neighbors and interest neighbors, such that both the higher-order social structure and higher-order interest network are directly injected in the user modeling process. Furthermore, we designed a multi-level attention network to attentively aggregate the graph and node level representations for better user modeling. Experimental results on two real-world datasets clearly showed the effectiveness of our proposed model. In the future, we would like to explore the graph reasoning models to explain the paths for users' behaviors.


\section*{Acknowledgements}

This work was supported in part by grants from  the National Natural Science Foundation of China( Grant No.  U19A2079, U1936219, 91846201, 61725203, 61732008, 61722204, 61932009), and the Foundation of Key Laboratory of Cognitive Intelligence, iFLYTEK, P.R., Chia(Grant No. COGOS-20190002), and  CAAI-Huawei
MindSpore Open Fund.

\begin{small}
\bibliographystyle{abbrv}
\bibliography{socialgcn_abbr_tkde_cl}

\begin{thebibliography}{10}

\bibitem{adomavicius2005toward}
G.~Adomavicius and A.~Tuzhilin.
\newblock Toward the next generation of recommender systems: A survey of the
  state-of-the-art and possible extensions.
\newblock {\em TKDE}, (6):734--749, 2005.

\bibitem{KDD2008influence}
A.~Anagnostopoulos, R.~Kumar, and M.~Mahdian.
\newblock Influence and correlation in social networks.
\newblock In {\em SIGKDD}, pages 7--15. ACM, 2008.

\bibitem{bahdanau2014neural}
D.~Bahdanau, K.~Cho, and Y.~Bengio.
\newblock Neural machine translation by jointly learning to align and
  translate.
\newblock In {\em ICLR}, 2015.

\bibitem{arXiv2017gcmc}
R.~v.~d. Berg, T.~N. Kipf, and M.~Welling.
\newblock Graph convolutional matrix completion.
\newblock In {\em SIGKDD}, 2018.

\bibitem{bruna2013spectral}
J.~Bruna, W.~Zaremba, A.~Szlam, and Y.~LeCun.
\newblock Spectral networks and locally connected networks on graphs.
\newblock In {\em ICLR}, 2014.

\bibitem{SIGIR2019efficient}
C.~Chen, M.~Zhang, C.~Wang, W.~Ma, M.~Li, Y.~Liu, and S.~Ma.
\newblock An efficient adaptive transfer neural network for social-aware
  recommendation.
\newblock In {\em SIGIR}, pages 225--234, 2019.

\bibitem{chen2020revisiting}
L.~Chen, L.~Wu, R.~Hong, K.~Zhang, and M.~Wang.
\newblock Revisiting graph based collaborative filtering: A linear residual
  graph convolutional network approach.
\newblock In {\em AAAI}, volume~34, pages 27--34, 2020.

\bibitem{defferrard2016convolutional}
M.~Defferrard, X.~Bresson, and P.~Vandergheynst.
\newblock Convolutional neural networks on graphs with fast localized spectral
  filtering.
\newblock In {\em NIPS}, pages 3844--3852, 2016.

\bibitem{deshpande2004item}
M.~Deshpande and G.~Karypis.
\newblock Item-based top-n recommendation algorithms.
\newblock {\em TOIS}, 22(1):143--177, 2004.

\bibitem{fan2019graph}
W.~Fan, Y.~Ma, Q.~Li, Y.~He, E.~Zhao, J.~Tang, and D.~Yin.
\newblock Graph neural networks for social recommendation.
\newblock In {\em WWW}, pages 417--426, 2019.

\bibitem{fan2019deep}
W.~Fan, Y.~Ma, D.~Yin, J.~Wang, J.~Tang, and Q.~Li.
\newblock Deep social collaborative filtering.
\newblock In {\em Recsys}, pages 305--313, 2019.

\bibitem{friedkin2006structural}
N.~E. Friedkin.
\newblock {\em A structural theory of social influence}, volume~13.
\newblock Cambridge University Press, 2006.

\bibitem{gong2018adaptive}
L.~Gong and Q.~Cheng.
\newblock Adaptive edge features guided graph attention networks.
\newblock {\em ArXiv}, abs/1809.02709, 2018.

\bibitem{trustsvd}
G.~Guo, J.~Zhang, and N.~Yorke-Smith.
\newblock Trustsvd: collaborative filtering with both the explicit and implicit
  influence of user trust and of item ratings.
\newblock In {\em AAAI}, pages 123--129, 2015.

\bibitem{TKDE2016novel}
G.~Guo, J.~Zhang, and N.~Yorke-Smith.
\newblock A novel recommendation model regularized with user trust and item
  ratings.
\newblock {\em TKDE}, 28(7):1607--1620, 2016.

\bibitem{he2018nais}
X.~He, Z.~He, J.~Song, Z.~Liu, Y.-G. Jiang, and T.-S. Chua.
\newblock Nais: Neural attentive item similarity model for recommendation.
\newblock {\em TKDE}, 30(12):2354--2366, 2018.

\bibitem{nmf}
X.~He, L.~Liao, H.~Zhang, L.~Nie, X.~Hu, and T.-S. Chua.
\newblock Neural collaborative filtering.
\newblock In {\em WWW}, pages 173--182, 2017.

\bibitem{itti1998model}
L.~Itti, C.~Koch, and E.~Niebur.
\newblock A model of saliency-based visual attention for rapid scene analysis.
\newblock {\em TPAMI}, (11):1254--1259, 1998.

\bibitem{socialmf}
M.~Jamali and M.~Ester.
\newblock A matrix factorization technique with trust propagation for
  recommendation in social networks.
\newblock In {\em Recsys}, pages 135--142, 2010.

\bibitem{jiang2012social}
M.~Jiang, P.~Cui, R.~Liu, Q.~Yang, F.~Wang, W.~Zhu, and S.~Yang.
\newblock Social contextual recommendation.
\newblock In {\em CIKM}, pages 45--54, 2012.

\bibitem{tkde2014scalable}
M.~Jiang, P.~Cui, F.~Wang, W.~Zhu, and S.~Yang.
\newblock Scalable recommendation with social contextual information.
\newblock {\em TKDE}, 26(11):2789--2802, 2014.

\bibitem{kipf2016semi}
T.~N. Kipf and M.~Welling.
\newblock Semi-supervised classification with graph convolutional networks.
\newblock In {\em ICLR}, 2017.

\bibitem{koren2008factorization}
Y.~Koren.
\newblock Factorization meets the neighborhood: a multifaceted collaborative
  filtering model.
\newblock In {\em SIGKDD}, pages 426--434, 2008.

\bibitem{PNAS2014experimental}
A.~D. Kramer, J.~E. Guillory, and J.~T. Hancock.
\newblock Experimental evidence of massive-scale emotional contagion through
  social networks.
\newblock {\em PNAS}, 111(24):8788--8790, 2014.

\bibitem{PNAS2012social}
K.~Lewis, M.~Gonzalez, and J.~Kaufman.
\newblock Social selection and peer influence in an online social network.
\newblock {\em PNAS}, 109(1):68--72, 2012.

\bibitem{li2015overlapping}
H.~Li, D.~Wu, W.~Tang, and N.~Mamoulis.
\newblock Overlapping community regularization for rating prediction in social
  recommender systems.
\newblock In {\em Recsys}, pages 27--34, 2015.

\bibitem{li2018deeper}
Q.~Li, Z.~Han, and X.-M. Wu.
\newblock Deeper insights into graph convolutional networks for semi-supervised
  learning.
\newblock In {\em AAAI}, pages 3538--3545, 2018.

\bibitem{kdd2018xdeepfm}
J.~Lian, X.~Zhou, F.~Zhang, Z.~Chen, X.~Xie, and G.~Sun.
\newblock xdeepfm: Combining explicit and implicit feature interactions for
  recommender systems.
\newblock In {\em SIGKDD}, pages 1754--1763, 2018.

\bibitem{liu2014influence}
Q.~Liu, B.~Xiang, E.~Chen, H.~Xiong, F.~Tang, and J.~X. Yu.
\newblock Influence maximization over large-scale social networks: A bounded
  linear approach.
\newblock In {\em CIKM}, pages 171--180, 2014.

\bibitem{WSDM2011recommender}
H.~Ma, D.~Zhou, C.~Liu, M.~R. Lyu, and I.~King.
\newblock Recommender systems with social regularization.
\newblock In {\em WSDM}, pages 287--296, 2011.

\bibitem{massa2007trust}
P.~Massa and P.~Avesani.
\newblock Trust-aware recommender systems.
\newblock In {\em Recsys}, pages 17--24, 2007.

\bibitem{pmf}
A.~Mnih and R.~R. Salakhutdinov.
\newblock Probabilistic matrix factorization.
\newblock In {\em NIPS}, pages 1257--1264, 2008.

\bibitem{monti2017geometric}
F.~Monti, M.~Bronstein, and X.~Bresson.
\newblock Geometric matrix completion with recurrent multi-graph neural
  networks.
\newblock In {\em NIPS}, pages 3697--3707, 2017.

\bibitem{tkde2014personalized}
X.~Qian, H.~Feng, G.~Zhao, and T.~Mei.
\newblock Personalized recommendation combining user interest and social
  circle.
\newblock {\em TKDE}, 26(7):1763--1777, 2014.

\bibitem{qiu2018deepinf}
J.~Qiu, J.~Tang, H.~Ma, Y.~Dong, K.~Wang, and J.~Tang.
\newblock Deepinf: Social influence prediction with deep learning.
\newblock In {\em SIGKDD}, pages 2110--2119, 2018.

\bibitem{rendle2010factorization}
S.~Rendle.
\newblock Factorization machines.
\newblock In {\em ICDM}, pages 995--1000, 2010.

\bibitem{bpr}
S.~Rendle, C.~Freudenthaler, Z.~Gantner, and L.~Schmidt-Thieme.
\newblock Bpr: Bayesian personalized ranking from implicit feedback.
\newblock In {\em UAI}, pages 452--461, 2009.

\bibitem{sun2018attentive}
P.~Sun, L.~Wu, and M.~Wang.
\newblock Attentive recurrent social recommendation.
\newblock In {\em SIGIR}, pages 185--194, 2018.

\bibitem{tang2013exploiting}
J.~Tang, X.~Hu, H.~Gao, and H.~Liu.
\newblock Exploiting local and global social context for recommendation.
\newblock In {\em IJCAI}, pages 2712--2718, 2013.

\bibitem{velivckovic2017graph}
P.~Veli{\v{c}}kovi{\'c}, G.~Cucurull, A.~Casanova, A.~Romero, P.~Lio, and
  Y.~Bengio.
\newblock Graph attention networks.
\newblock In {\em ICLR}, 2018.

\bibitem{wang2019neural}
X.~Wang, X.~He, M.~Wang, F.~Feng, and T.-S. Chua.
\newblock Neural graph collaborative filtering.
\newblock In {\em SIGIR}, pages 165--174, 2019.

\bibitem{wu2019hierarchical}
L.~Wu, L.~Chen, R.~Hong, Y.~Fu, X.~Xie, and M.~Wang.
\newblock A hierarchical attention model for social contextual image
  recommendation.
\newblock {\em TKDE}, 32(10):1754--1867, 2020.

\bibitem{leDiffnet}
L.~Wu, P.~Sun, Y.~Fu, R.~Hong, X.~Wang, and M.~Wang.
\newblock A neural influence diffusion model for social recommendation.
\newblock In {\em SIGIR}, pages 235--244, 2019.

\bibitem{wu2018collaborative}
L.~Wu, P.~Sun, R.~Hong, Y.~Ge, and M.~Wang.
\newblock Collaborative neural social recommendation.
\newblock {\em TSMC: Systems}, pages 1--13, 2018.

\bibitem{wu2019dual}
Q.~Wu, H.~Zhang, X.~Gao, P.~He, P.~Weng, H.~Gao, and G.~Chen.
\newblock Dual graph attention networks for deep latent representation of
  multifaceted social effects in recommender systems.
\newblock In {\em WWW}, pages 2091--2102, 2019.

\bibitem{xu2015show}
K.~Xu, J.~Ba, R.~Kiros, K.~Cho, A.~Courville, R.~Salakhudinov, R.~Zemel, and
  Y.~Bengio.
\newblock Show, attend and tell: Neural image caption generation with visual
  attention.
\newblock In {\em ICML}, pages 2048--2057, 2015.

\bibitem{ICLR2019powerful}
K.~Xu, W.~Hu, J.~Leskovec, and S.~Jegelka.
\newblock How powerful are graph neural networks?
\newblock In {\em ICLR}, 2019.

\bibitem{ying2018graph}
R.~Ying, R.~He, K.~Chen, P.~Eksombatchai, W.~L. Hamilton, and J.~Leskovec.
\newblock Graph convolutional neural networks for web-scale recommender
  systems.
\newblock In {\em SIGKDD}, pages 974--983, 2018.

\bibitem{ijcai2019star-gcn}
J.~Zhang, X.~Shi, S.~Zhao, and I.~King.
\newblock {STAR-GCN:} stacked and reconstructed graph convolutional networks
  for recommender systems.
\newblock In {\em IJCAI}, pages 4264--4270, 2019.

\bibitem{zhao2014leveraging}
T.~Zhao, J.~McAuley, and I.~King.
\newblock Leveraging social connections to improve personalized ranking for
  collaborative filtering.
\newblock In {\em CIKM}, pages 261--270, 2014.

\bibitem{zheng2018spectral}
L.~Zheng, C.-T. Lu, F.~Jiang, J.~Zhang, and P.~S. Yu.
\newblock Spectral collaborative filtering.
\newblock In {\em Recsys}, pages 311--319, 2018.

\bibitem{zhou2018graph}
J.~Zhou, G.~Cui, Z.~Zhang, C.~Yang, Z.~Liu, L.~Wang, C.~Li, and M.~Sun.
\newblock Graph neural networks: A review of methods and applications.
\newblock {\em arXiv preprint arXiv:1812.08434}, 2018.

\end{thebibliography}
\end{small}

\begin{IEEEbiography}[{\includegraphics[width=1in,height=1.25in,clip,keepaspectratio]{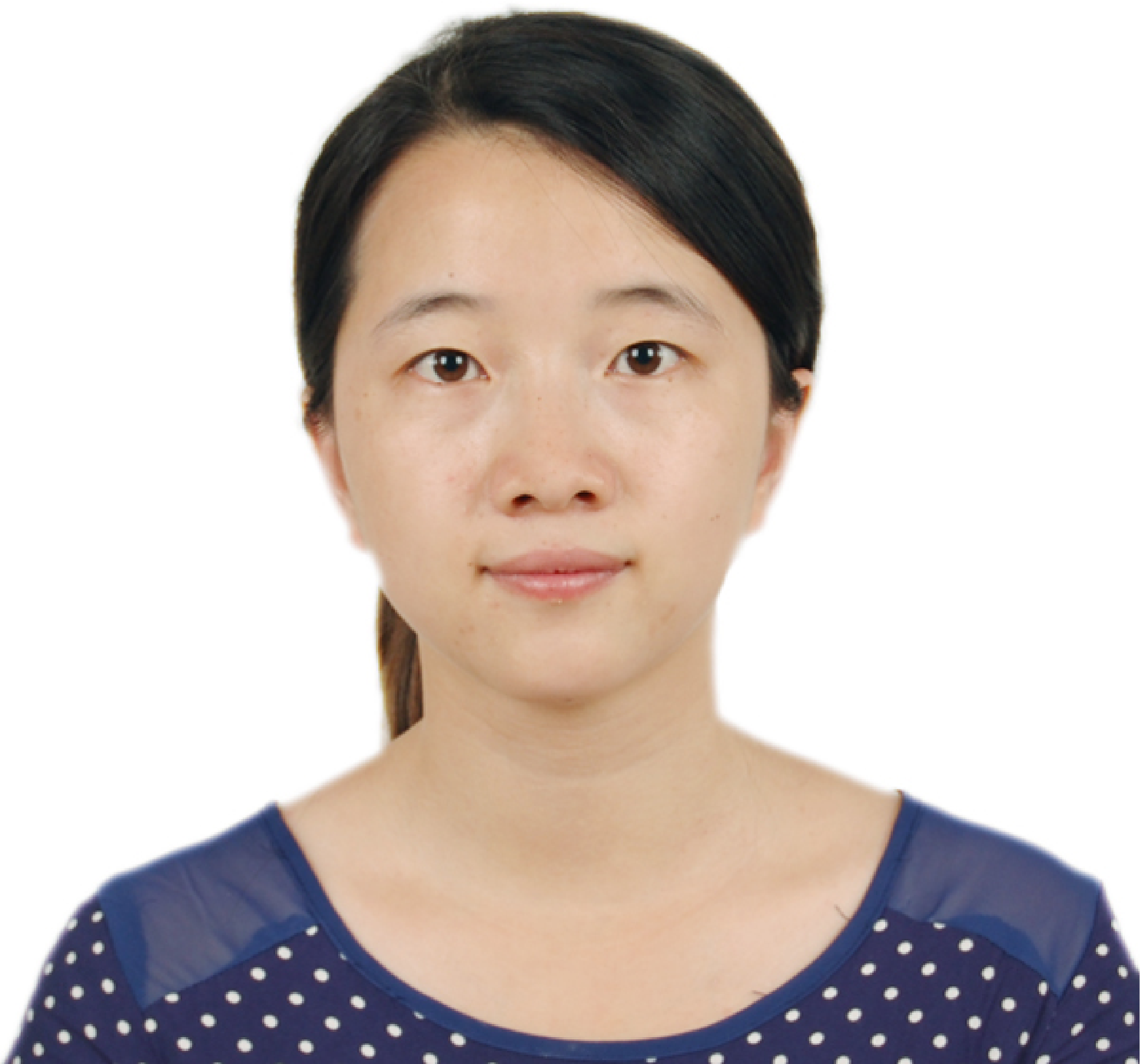}}]
{Le Wu} is currently an associate professor at the Hefei University of Technology (HFUT), China. She received the Ph.D. degree from the University of Science and Technology of China (USTC). Her general area of research  interests is data mining, recommender systems and social network analysis. She has published more than 40 papers in referred journals and conferences. Dr. Le Wu is the recipient of the Best of SDM 2015 Award, and the Distinguished Dissertation Award from China Association for Artificial Intelligence (CAAI) 2017.
\end{IEEEbiography}

\begin{IEEEbiography}[{\includegraphics[width=1in,height=1.25in,clip,keepaspectratio]{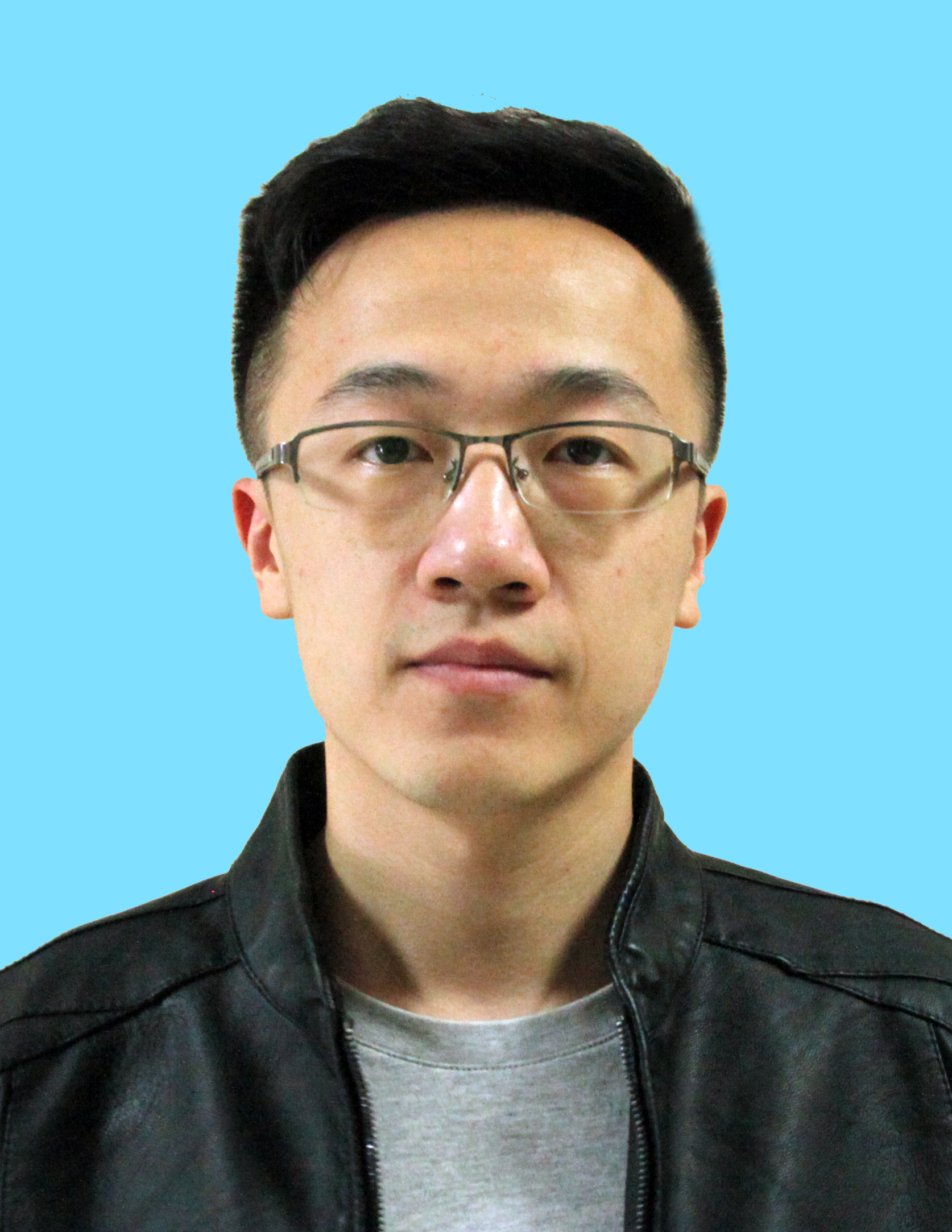}}]
{Junwei Li} is currently working towards the Ph.D. degree at Hefei University of Technology, China. His research interests include data mining and recommender systems.
\end{IEEEbiography}

\begin{IEEEbiography}[{\includegraphics[width=1in,height=1.25in,clip,keepaspectratio]{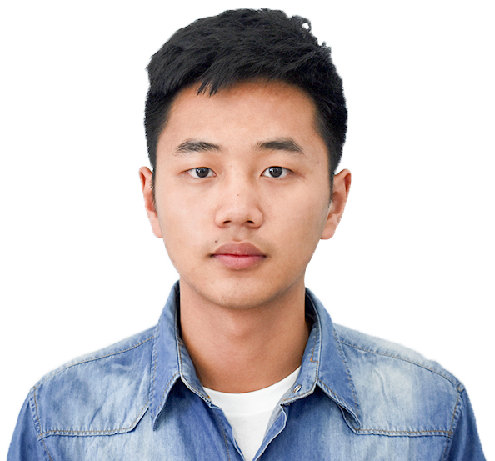}}]
{Peijie Sun} is a Ph.D. student with the Hefei University of Technology. He received the master degree from the same university
in 2018. He has published several papers in leading conferences and journals, including SIGIR and IEEE Trans. on SMC: Systems. His current research interests include data mining and recommender systems.
\end{IEEEbiography}

\begin{IEEEbiography}[{\includegraphics[width=1.0in,height=1.25in,clip,keepaspectratio]{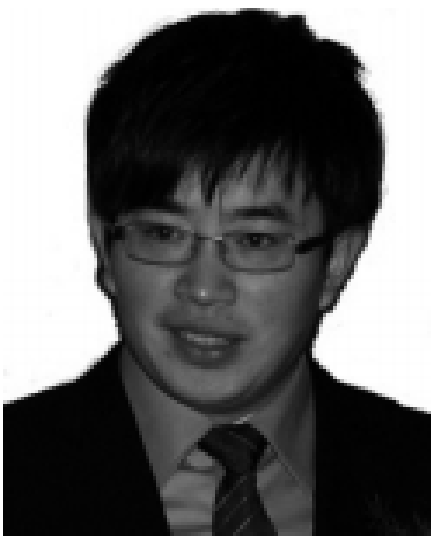}}]
{Richang Hong} is currently a professor at HFUT. He  received the Ph.D. degree from USTC, in 2008. He has co-authored over 60 publications in the areas of his research interests, which include multimedia question answering, video content analysis, and pattern recognition. He is a member of the Association for Computing Machinery. He was a recipient of the best paper award in the ACM Multimedia 2010.
\end{IEEEbiography}

\begin{IEEEbiography}[{\includegraphics[width=1in,height=1.25in,clip,keepaspectratio]{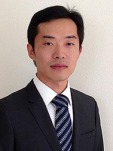}}]
{Yong Ge} is an assistant professor of Management Information Systems in University of Arizona. He received the Ph.D. degree in information technology from Rutgers, The State University of New Jersey in 2013. His research interests include data mining and business analytics. He received the ICDM-2011 Best Research Paper Award.. He has published prolifically in refereed journals and conference proceedings, such as IEEE Transactions on Knowledge and Data Engineering, ACM Transactions on Information Systems, ACM Transactions on Knowledge Discovery from Data, ACM SIGKDD, SIAM SDM, IEEE ICDM, and ACM RecSys. He also was the Program Committee member at ACM SIGKDD, IEEE ICDM etc.
\end{IEEEbiography}

\begin{IEEEbiography}[{\includegraphics[width=1.0in,height=1.25in,clip,keepaspectratio]{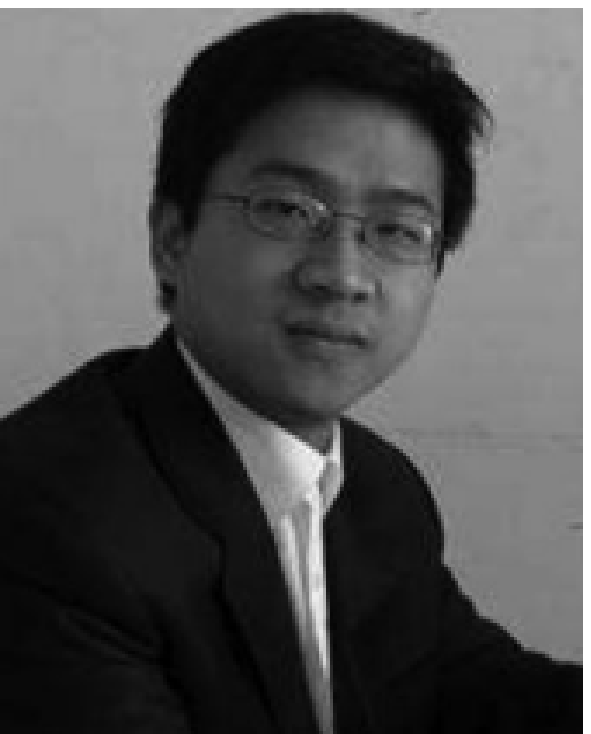}}]
{Meng Wang} is a professor at the Hefei University of Technology,
China. He received his B.E. degree and Ph.D. degree in the Special
Class for the Gifted Young and the Department of Electronic
Engineering and Information Science from the University of Science and
Technology of China (USTC), Hefei, China, in 2003 and 2008,
respectively. His current research interests include multimedia
content analysis, computer vision, and pattern recognition. He has
authored more than 200 book chapters, journal and conference papers in
these areas. He is the recipient of the ACM SIGMM Rising Star Award 2014.
He is an associate editor of IEEE Transactions on Knowledge and Data
Engineering (IEEE TKDE), IEEE Transactions on Circuits and Systems
for Video Technology (IEEE TCSVT), IEEE Transactions on Multimedia (IEEE TMM), and IEEE Transactions on Neural Networks and Learning Systems (IEEE TNNLS).
\end{IEEEbiography}

\end{document}